\documentclass[11pt]{article}
\usepackage{amssymb,amscd,amsmath,amsfonts}
\usepackage{amsthm,latexsym,multibox,feynmf}

\usepackage{hyperref}

\def\be{\begin{equation}}
\def\ee{\end{equation}}
\def\ba{\begin{array}}
\def\ea{\end{array}}

\def\ft#1#2{{\textstyle{\frac{\scriptstyle #1}{\scriptstyle #2}}}}

\def\be{\begin{equation}}
\def\ee{\end{equation}}
\def\ba{\begin{array}}
\def\ea{\end{array}}

\def\l{\lambda}

\newlength{\blength}
\renewcommand{\proof}[1]{\vspace{-.05cm}
\begin{list}{\bf Proof:}
{\listparindent=\parindent\parsep=0pt \labelwidth=-0.5cm
\labelsep=\parindent \addtolength{\labelsep}{-\blength}
\addtolength{\labelsep}{1.5cm} \itemindent=-\blength
\addtolength{\itemindent}{\parindent} \leftmargin=1.0cm} \item
#1~$\qedsymbol$\end{list} \vspace{.0cm}}

\usepackage{color}
\definecolor{rougef}{rgb}{0.56,0,0}
\definecolor{vertf}{rgb}{0,0.5,0}
\definecolor{bleuf}{rgb}{0,0,0.8}
\definecolor{violetf}{rgb}{0.5,0,0.5}

\begin{document}

\begin{center}
{\textbf{
{\LARGE How higher-spin gravity surpasses the spin two barrier:}\\[11pt]
{\LARGE no-go theorems versus yes-go examples} \\}}
\end{center}
\vspace*{1cm}

\begin{center}
{Xavier Bekaert$^{\clubsuit,}$\footnote{E-mail address: 
\tt{xavier.bekaert@lmpt.univ-tours.fr}},
Nicolas Boulanger$^{\spadesuit,}$\footnote{Research Associate of the Fund for 
Scientific Research-FNRS (Belgium); \tt{nicolas.boulanger@umons.ac.be}}
and Per Sundell$^{\spadesuit,}$\footnote{Ulysse Incentive Grant for Mobility in 
Scientific Research, F.R.S.-FNRS; \tt{per.sundell@umons.ac.be}} }
\end{center}

\vspace*{1cm}

\begin{footnotesize} 
\begin{center}
$^\clubsuit$
Laboratoire de Math\'ematiques et Physique Th\'eorique\\
Unit\'e Mixte de Recherche $6083$ du CNRS\\
F\'ed\'eration de Recherche $2964$ Denis Poisson\\
Universit\'e Fran\c{c}ois Rabelais, Parc de Grandmont\\
37200 Tours, France \\
\vspace{2mm}{\tt \footnotesize }
$^{\spadesuit}$
Service de M\'ecanique et Gravitation\\
Universit\'e de Mons -- UMONS\\
20 Place du Parc\\
7000 Mons, Belgique\\
\vspace*{.3cm}
\end{center}
\end{footnotesize}
\vspace*{1cm}

\begin{abstract}

Aiming at non-experts, we explain the key mechanisms of higher-spin extensions of 
ordinary gravities in four dimensions and higher. We first overview various no-go theorems for low-energy 
scattering of massless particles in flat spacetime. In doing so we dress a 
dictionary between the S-matrix and the Lagrangian approaches, exhibiting their 
relative advantages and weaknesses, after which we high-light potential loop-holes 
for non-trivial massless dynamics.
We then review positive yes-go results for non-abelian cubic higher-derivative 
vertices in constantly curved backgrounds. Finally we outline how higher-spin 
symmetry can be reconciled with the equivalence principle in the presence of a 
cosmological constant leading to the Fradkin--Vasiliev vertices and Vasiliev's 
higher-spin gravity with its double perturbative expansion (in terms of numbers of 
fields and derivatives).

\end{abstract}

\newpage

{\small{
\tableofcontents\newpage}}

\section{\large Introduction}

This review is an attempt at a non-technical summary of how higher-spin 
gravity\footnote{By the terminology ``higher-spin gravity" we mean a theory where 
an extension of the spacetime isometry algebra by higher-spin generators is 
gauged.} 
manages to surpass the spin-two barrier: the stringent constraints on low-energy 
scattering in flat spacetime that seemingly forbid massless particles with spins 
greater than two to participate in the formation of any interacting quantum field 
theory.\footnote{These constraints on massless particle scattering only appear in spacetimes 
of dimension $D\geqslant 4$ to 
which we shall restrict our attention in the present paper. 
Indeed, in dimension $D\leqslant 3$ massless fields of helicity $s\geqslant 2$ have 
no local propagating degrees of freedom. Pure massless higher-spin gravities in lower dimensions are of Chern-Simons type which do not 
share most of the exotic features of their higher-dimensional cousins discussed here.} 
While this may seem to call for radical measures, there exists a 
relatively conservative yet viable way out, namely the dual usage of the 
cosmological constant as critical mass (infrared cutoff) and dimensionful  
coupling constant. 
This dual-purpose treatment of the 
cosmological constant leads to a successful exchange of what are leading and 
sub-leading terms in minimal coupling that lifts the spell of the no-go 
theorems --- and in particular reconciles higher-spin gauge symmetry with the 
equivalence principle --- leading up to the Fradkin--Vasiliev cubic 
action~\cite{Fradkin:1987ks,Fradkin:1986qy,Vasiliev:2001wa,Alkalaev:2002rq,Vasiliev:2011xf} 
and Vasiliev's fully nonlinear equations of motion\footnote{The precise link 
between, on the one hand, the Fradkin--Vasiliev cubic action and, on the other 
hand, the fully interacting Vasiliev equations, remains to be found.}
\cite{Vasiliev:1990en,Vasiliev:1992av,Vasiliev:2003ev} 
(see e.g. \cite{Vasiliev:2004qz,Vasiliev:2004cp,Bekaert:2005vh} 
for some reviews).

Since our aim is to outline main ideas and results, we shall refrain from being 
technical and refer the reader to the already existing literature whenever 
necessary. Moreover, 
we shall mostly stick throughout the body of the paper to the Fronsdal programme~\cite{Fronsdal:1978rb}, 
\emph{i.e.} the standard perturbative off-shell implementation of non-abelian gauge 
deformations starting from the Fronsdal actions in constantly curved 
backgrounds. It is the gauge algebra (not necessarily an internal algebra) 
that we require to become non-abelian like the diffeomorphism 
algebra in Einstein gravity.
As for Vasiliev's higher-spin gravity --- presently the most 
far-reaching construction of a full higher-spin gauge theory albeit so far only 
known on-shell --- we shall restrict ourselves\footnote{We shall thus leave out 
many other of the interesting features of the Vasiliev system, such as its 
unfolded, or Cartan integrable, formulation, and the link between its
 first-quantization, deformed Wigner oscillators, singletons and compositeness of 
massless particles in anti-de Sitter spacetime.} to a more brief address of how it 
presents a natural framework for a string-theory-like double perturbative expansion.

Now, why are higher-spin gauge fields interesting? Although massless fields of 
spin greater than two make perfect sense at the free level, their quantum 
interactions pose a main challenge to modern theoretical physics. In a nut-shell, 
the problematics can be summarized as follows: consistent non-abelian 
higher-spin gauge symmetries induce local higher-derivative generalizations of 
translations that seem to call for a non-trivial 
bosonic extension of spacetime itself, thus interfering with the basic assumptions 
of canonical second-quantization that led up to the notion of free fields to begin 
with. Thus a satisfactory resolution seems certainly much more demanding than even 
that of quantizing ordinary general relativity (though the prolongation of 
the Einstein--Cartan reformulation of general relativity as a soldered Yang--Mills theory for the 
space-time isometry algebra soon leads to infinite-dimensional algebras as well) which 
actually leaves room for a naive optimism: the quantization of higher-spin gauge 
theories could lead to a radically new view on quantum field theory altogether, 
and in particular on the formidable spin-two barrier set up by the requirement of 
power-counting renormalizability.

Indeed, at the classical level, there exist the aforementioned higher-spin 
gravities 
\cite{Vasiliev:1990en,Vasiliev:1992av,Vasiliev:1995dn,Sezgin:1998gg,Sezgin:2001zs,Sezgin:2001yf,Sezgin:2002ru,Vasiliev:2003ev}: 
these are special instances of interacting higher-spin gauge theories constituting 
what one may think of as the simplest possible higher-spin extensions of general relativity. 
Their minimal bosonic versions (in $D\geqslant 4$
ordinary space-time dimensions) consist of a propagating scalar, metric and tower 
of massless fields of even 
spins, $s=4, 6, \ldots$
(these models can then be extended by various forms of ``matter'' 
and suitable higher-spin counterparts --- 
in a supersymmetric set-up in case fermions are included).

As already mentioned, a key feature of higher-spin gravity is its double 
perturbative expansion: besides the expansion in numbers of fields, weighted 
by a dimensionless coupling $g\,$, there is a parallel albeit strongly coupled
expansion in numbers of pairs of derivatives, 
weighted by a dimensionful parameter, 
the cosmological constant $\Lambda\,$, thus serving both as infrared 
and ultraviolet cutoff.
Hence classical higher-spin gravity prefers a non-vanishing cosmological 
constant --- unlike string theory in flat spacetime which also has a double 
perturbative expansion but with a strictly massless sector accessible at low 
energies in a weakly coupled derivative expansion.

Taking higher-spin gravity seriously as a model for quantum gravity, the key issue 
is thus whether its loop corrections\footnote{For related issues within the 
AdS/CFT correspondence, see \cite{Sezgin:2002rt,Klebanov:2002ja} 
and the recent advances \cite{Giombi:2009wh,Giombi:2010vg} due to 
Giombi and Yin, which altogether point to that four-dimensional higher-spin 
gravity should have a surprisingly simple
ultraviolet behavior as a quantum field theory in anti-de Sitter spacetime, in 
the sense that its boundary dual is weakly coupled or even free, with a simple 1/N-expansion.} 
--- which are given in a weak-field expansion more reminiscent of  
the perturbative expansion of string theory than that of general relativity --- 
may generate masses dynamically for the higher-spin fields? 
Remarkably, relying on arguments based on the Anti de Sitter/Conformal Field Theory 
(AdS/CFT) correspondence 
\cite{Girardello:2002pp}, the answer seems affirmative: the pattern of symmetry 
breaking is similar in spirit to that of ordinary Quantum Chromo Dynamics (QCD), 
with spin playing the r\^ole of color, the metric playing the r\^ole of an abelian 
gauge field, 
and the Goldstone modes being two-particle states; in the leading order in perturbation 
theory, the spin-$s$ field acquires mass for $s>2$ while the spin $s-1$ 
Goldstone mode is the lightest bound state (in its parity sector) between the physical 
scalar and the massless spin $s-2$ particle. 
The crucial missing ingredient is a ``confinement mechanism'' that would cause 
$g$ to become large at low enough energies, thus creating a mass-gap leading to a 
low-energy effective quantum gravity.

Thus, the quantization of higher-spin gauge theories may lead to interesting 
models providing deepened insights into the interplay between quantum mechanics 
and geometry. These might be of relevance not only in the high-energy limit of 
quantum gravity and string theory, but also for providing new ideas in 
observational physics, such as for example in cosmology, where weakly coupled 
massless particles could serve as dark matter candidates. Finally, the development 
of the quantum theory of higher-spin fields may service as a source of inspiration 
for seeking and testing new methods in quantum field theory, such as the 
application of deformation and geometric quantizations as well as 
topological models to dynamical systems with local degrees of freedom.

Having provided all of these motivations for quantizing higher-spin gauge fields, 
it is perhaps surprising to discover that there is drastic gap between Vasiliev's
on-shell approach to higher-spin gravity based on gauging a non-abelian 
global symmetry algebra and the Fronsdal programme: the latter 
has so far only been partially completed, mainly at the cubic level (for a recent 
discussion on this issue, see e.g. \cite{Bengtsson:2008mw} and references 
therein). 
Hence a key 
question\footnote{Here we wish to stress that it is only by closing the quartic 
order that the cubic Lagrangian -- including cubic curvature couplings known as 
cubic Born--Infeld terms --- will be completely fixed (if it exists). Due to the 
double perturbative expansion, the Born--Infeld couplings dominate over the 
minimal couplings in physical amplitudes (assuming a deformed Fronsdal action with 
finite Born--Infeld ``tail'') and hence the quartic-closure problem must be addressed 
prior to any attempts to do physics with incomplete cubic actions. 
In other words, analyses based solely on current exchange may receive 
large corrections due to the exotic usage of the cosmological constant.} is whether 
the Fronsdal programme can be completed at the quartic level, 
even in the case of the aforementioned minimal bosonic model? 
This apparently straightforward problem may keep a number of interesting surprises 
in store --- in particular in view of the aforementioned properties of the AdS/CFT 
correspondence \cite{Sezgin:2002rt,Giombi:2009wh,Giombi:2010vg} which have been 
derived using a rather different approach --- as we shall return to in Section \ref{Sec:VE} 
and summarize in the Conclusions (Section \ref{Sec:conclusions}).

As far as more general interacting quantum field theories with higher-spin fields are concerned, 
open string field theory in flat spacetime provides a basic example thereof albeit with massless 
sector restricted to spins less than or equal to one. Recently, motivated by the similarities 
between open string theory and higher-spin gravities mainly at the level free 
fields \cite{Francia:2002pt,Francia:2006hp}, Sagnotti and Taronna \cite{Sagnotti:2010at} 
have deconstructed its first Regge trajectory and arrived at the germs of the non-abelian 
interactions for massless totally symmetric tensors in flat spacetime 
\cite{Boulanger:2006gr,Boulanger:2008tg} whose deformations into (A)dS spacetimes 
\cite{Boulanger:2008tg} lead to the Fradkin--Vasiliev cubic vertices. 
Moreover, in \cite{Polyakov:2009pk} D. Polyakov has proposed to extend the open superstring 
in flat spacetime by sectors of states with novel world-sheet ghost numbers containing massless 
higher-spin particles in interaction. He has also managed to show \cite{Polyakov:2010qs} 
that these higher-spin states interact with the closed-string graviton and that these interaction 
reproduce the aforementioned germs of \cite{Boulanger:2006gr,Boulanger:2008tg}. 

As far as actual tensionless limits of strings are concerned, there is a vast literature which we 
cannot cover here. Of the various results that have been obtained, we simply wish to point to 
the rather drastic difference between tensionless limits of, on the one hand, the open string 
in flat space and, on the other hand, the closed string in anti-de Sitter spacetime. 
A precise version of the former was taken in \cite{Bonelli:2003kh,Buchbinder:2006eq,Fotopoulos:2007nm,Fotopoulos:2007yq}. It yields deformed Fronsdal actions albeit with abelian p-form-like vertices that do not contain the non-abelian interactions characteristic of the higher-spin gravities to be discussed in this review. Whether there exists a refined limit in spirit of the aforementioned deconstruction in \cite{Sagnotti:2010at}, leading to such couplings, remains to be seen. 

As far as the closed AdS string is concerned, it exhibits a novel physical phenomenon 
that has no flat-space analog whereby solitons, carrying quantum numbers of singletons, 
are formed at cusps \cite{Engquist:2005yt}; in the tensionless limit, their dynamics can 
be extracted by discretizing the Nambu-Goto action and degenerating spacetime to the Dirac 
hypercone leading to a direct connection between Vasiliev's higher-spin gravities and 
tensionless closed strings in which the graviton on both sides is identified \cite{Engquist:2005yt}. 
The resulting physical picture is also in accordance with the holographic proposals 
in \cite{Sundborg:2000wp,Sezgin:2002rt} later dubbed ``la grande bouffe'' \cite{Bianchi:2003wx}.

Although these string-related theories are extremely interesting in their own right, 
in this paper we shall mainly be concerned with non-abelian 
interactions for strictly massless fields in flat spacetime and for 
their (A)dS analogs with their critical masses and the related higher-spin gravity.

In the case of strictly massless fields in flat spacetime, 
many $S$-matrix no-go theorems can be found in the literature 
\cite{Weinberg:1964ew,Grisaru:1976vm,Coleman:1967ad,Haag:1974qh,Benincasa:2007xk,Porrati:2008rm,Benincasa:2011pg} 
that seemingly forbid interacting massless higher-spin particles. 
Since the relative strength 
of no-go theorems is measured by the weakness of their hypotheses, the $S$-matrix 
approach is usually advertised because it does not require assumptions about locality 
nor the Poincar\'e-covariant realization of the incoming quanta.
At a closer inspection, however, it turns out that the $S$-matrix no-go results
obtained so-far only concern the spin-$s$ couplings involving $s$ derivatives 
such as, for example, two-derivative couplings between the graviton and other fields.

If one accepts that the 
spin-$s$ couplings contain more than $s$ derivatives, 
then these $S$-matrix arguments need to be reconsidered, 
and since the higher-spin interaction problem presents itself already at the 
classical level, it is anyway more satisfactory to pursue this analysis starting from 
purely Lagrangian arguments. 
And indeed, numerous cubic vertices, consistent at this order, 
have been found over the years in Minkowski and (A)dS spacetimes. 
They all exhibit higher-derivative couplings and will be reviewed here, 
as well as their relations with the Fradkin--Vasiliev vertices.

In summary, it may prove to be useful to confront
the no-go theorems with the yes-go examples already in the classical Lagrangian 
framework, in order to emphasize the underlying 
assumptions of the no-go theorems, even if it may require an extra assumption 
about perturbative locality.

The paper is organized as follows: In Section \ref{Nogoreview}, we begin by spelling 
out 
the gauge principle in perturbative quantum field theory and its ``standard'' 
implementation within the Fronsdal programme for higher-spin gauge 
interactions. We then survey the problematics of non-trivial 
scattering of massless particles of spin greater than two in flat spacetime, and 
especially its direct conflict with the equivalence principle. 
In Section \ref{wayout}, we list possible ways to evade these negative results --- both 
within and without the Fronsdal programme. 
In Section \ref{yesgo} we review results where consistent higher-spin interactions 
have been found, both in flat and (A)dS spacetimes. Due to the fact that 
consistent interacting higher-spin gravities indeed exist, 
at least for gauge algebras which are infinite-dimensional extensions of the 
(A)dS isometry algebra, an important question is related to the possible symmetry 
breaking mechanisms that would give a mass to the higher-spin gauge fields. 
This is briefly discussed in Subsection \ref{break}. 
After reviewing why a classically complete theory is crucial in higher-spin 
gravity, we lay out in Section \ref{Sec:VE} 
the salient features of Vasiliev's approach to a class of potentially viable 
models of quantum gravity. We end our presentation with a few stringy remarks in 
Section \ref{sec:extended}. 
We conclude in Section \ref{Sec:conclusions} where we also summarize some interesting open 
problems. Finally we devote two Appendices to the review of some $S$-matrix 
no-go theorems and to their reformulation in Lagrangian language. More precisely, 
Appendix \ref{sec:Gra} focuses on Weinberg's low-energy theorem while 
Appendix \ref{sec:S} concentrates on the Weinberg--Witten theorem and its recent adaptation 
to gauge theories by Porrati.

\section{\large No-go theorems in flat spacetime}\label{Nogoreview}

This section presents various theorems\footnote{The $S$-matrix no-go 
theorem \cite{Benincasa:2007xk} is not discussed here because it relies on 
slightly stronger assumptions than the others --- see e.g. the conclusion 
of \cite{Porrati:2008rm} for more comments.} that constrain interactions between 
massless particles in flat spacetime --- potentially ruling out non-trivial 
quantum field theories with gauge fields with spin $s>2$ and vanishing 
cosmological constant. The aim is to scrutinize some of their hypotheses in order 
to exhibit a number of conceivable loop-holes that may lead to modified theories 
including massless higher-spin, as summarized in Subsection \ref{wayout}.
 
\subsection{Preamble: the gauge principle and the Fronsdal programme}\label{GP}

The key feature of the field-theoretic description of interacting massless 
particles is the \emph{gauge principle: a sensible 
perturbation theory requires compatibility between the interactions and some 
deformed version of the abelian gauge symmetries of the free limit}. The necessity 
of gauge invariance in perturbative quantum field theory stems from the fact that 
one and the same massless particle, thought of as a representation of the space-
time isometry group, in general admits (infinitely) many implementations in terms 
of quantum fields sitting in different Lorentz tensors obeying respective free 
equations of motion. For more information, see e.g. \cite{Skvortsov:2008vs,Boulanger:2008up}. 

Only a subset of these ``carriers'', namely the primary curvature tensors and all 
of their derivatives, actually transform tensorially under isometry (implemented 
quantum-mechanically via similarity transformations). The remaining carriers are 
different types of potentials obtained by integrating various curvature Bianchi 
identities (and which one may thus think of as representing different ``dual 
pictures'' of one and the same particle); such integrals in general transform 
under isometry with inhomogeneous pieces that one can identify as abelian gauge 
transformations. 

Thus, in the standard perturbative interaction picture one is led to the 
\emph{Fronsdal programme}: the construction of interaction Hamiltonians starting from 
Lorentz invariant and hence gauge invariant non-linear Lagrangians built from the 
aforementioned carriers.

We wish to stress that the Fronsdal programme is based on a working hypothesis: 
that standard canonical quantization of free fields in ordinary spacetime is 
actually compatible with the presence of higher-spin translations in higher-spin 
gauge theories. We shall proceed in this spirit in the bulk of this paper. 

\subsection{The Weinberg low-energy theorem}\label{lowenergy}

The Weinberg low-energy theorem is essentially a byproduct of dealing with the more general problem of emissions of soft massless particles. 
Consider a (non-trivial) scattering process involving $N$ external particles with (say, ingoing) momenta $p_i$ ($i=1,2,\ldots, N$) and spin $s_i\,$.
The emission of an additional massless particle of integer spin $s$ with arbitrary soft momentum by the $i$th external particle is controlled by a cubic vertex of type $s$-$s_i$-$s_i$ (\emph{i.e.} between a gauge boson of spin $s$ and two particles of spin $s_i$) with coupling constant $g^{(s)}_i$.
The Weinberg low-energy theorem \cite{Weinberg:1964ew} states that Lorentz 
invariance of (or equivalently, the absence of unphysical degrees of freedom from) the deformed amplitude imposes a conservation law of order $s-1$ on the $N$ external momenta:\footnote{For pedagogical reviews, see e.g. \cite{Weinberg:1995mt}, Section 13.1 or
\cite{Blagojevic:2002du}, Appendix G.}
\begin{equation}
\boxed{\;\sum_{i=1}^N g^{(s)}_i\,p_i^{\mu_1}\ldots p_i^{\mu_{s-1}}=0\;}\quad .
\label{lowen}
\end{equation}

\subsubsection{Charge conservation: the spin-one case}

\noindent Lorentz invariance for the emission of a soft massless spin-one particle 
(like a photon) leads to the conservation law $\sum_{i} g^{(1)}_i=0\,$;
thus it requires the conservation of the coupling constants (like the electric charges) 
that characterize the interactions of these particles at low energies.

In order to prepare the ground for further discussion, let us denote by ``electromagnetic minimal coupling'' the coupling of a charged 
particle to the electromagnetic field obtained by replacing the partial 
derivatives appearing in the Lagrangian describing the free, charged matter field 
in flat space, by the $u(1)$-covariant derivative, \emph{viz.} 
$\partial_{\mu} \rightarrow \partial_{\mu} - \,$i\,$ g^{(1)}_i A_{\mu}$.

\subsubsection{Equivalence principle: the spin-two case}\label{s2case}

\noindent As argued by Weinberg \cite{Weinberg:1964ew}, the equivalence principle 
can be recovered as the spin-two case of his low-energy theorem. 
On one side, Lorentz invariance for the emission of a soft massless spin-two 
particle leads to the conservation law $\sum_{i} g^{(2)}_i\,p_i^\mu=0\,$. 
On the other side, translation invariance implies momentum conservation 
$\sum_{i} p_i^\mu=0\,$. 
Therefore, for generic momenta, Poincar\'e invariance requires all coupling 
constants to be equal: $g_i^{(2)}=g_j^{(2)}=:g^{(2)}$ ($\forall\, i,j$). 
In other words, massless particles of spin-two must couple in the same 
way to all particles at low energies. 

This result has far-reaching consequences as it resonates with two deep properties 
of gravity, namely its uniqueness and its universality. 
On the one hand, the local theory of a self-interacting 
massless spin-two particle is essentially\footnote{See e.g. 
\cite{Boulanger:2000rq} 
for a precise statement of the very general hypotheses, and see refs therein for 
previous literature on this issue.} 
\textit{unique}: in the low-energy regime (at most two derivatives in the 
Lagrangian) it must be described by the Einstein--Hilbert action. 
Therefore, the massless spin-two particle rightfully deserves the name 
``graviton''\footnote{A thorough discussion on the observability of the graviton 
is presented in \cite{Rothman:2006fp,Boughn:2006st}.}. 
On the other hand, the gravitational interaction is also \emph{universal} 
\cite{Weinberg:1964ew}: 
if there exists a single particle that couples minimally to the graviton, then all particles coupled to at least one of them must also couple minimally to the graviton. 
According to Weinberg himself, this theorem is the 
expression of the equivalence principle in quantum field theory, so, from now 
on, it will be referred to as the Weinberg equivalence principle. 
A proper understanding of this crucial theorem involves, however, some subtleties 
on the precise meaning of ``minimal coupling''. 

Let us consider the quadratic Lagrangian 
${\cal L}^{(0)}(\varphi_s,\partial\varphi_s)$ describing a free spin-$s$ ``matter'' field 
denoted by $\varphi_s\,$.
In general relativity, the equivalence principle may be expressed by the Lorentz 
minimal coupling prescription, \textit{i.e.} the assumption that the 
transformation rules of tensor fields under the Poincar\'e group extend naturally 
to the diffeomorphism group and the replacement of partial 
derivatives by Lorentz-covariant ones, \emph{viz.} 
$\partial\rightarrow \nabla=\partial+g^{(2)}\Gamma_{\rm lin}+\cdots\,$, in 
the matter sector. 
It must be observed that this prescription does not apply to the spin-two field 
itself because the Einstein--Hilbert Lagrangian is \textit{not} the covariantization 
of the Fierz--Pauli quadratic Lagrangian 
${\cal L}^{(0)}(\varphi_2,\partial\varphi_2)\,$.

One focuses on cubic couplings ${\cal L}^{(1)}(h,\varphi_s,\partial\varphi_s)$ of 
the type $2$-$s$-$s$, \textit{i.e.} linear in the spin-two field $h_{\mu\nu}$ and 
quadratic in the spin-$s$ field $\varphi_s\,$. 
The symmetric tensor of rank two 
$\Theta^{\mu\nu}:=\delta{\cal L}^{(1)}/\delta h_{\mu\nu}$ 
is bilinear in the spin-$s$ field. 
For consistency with the linearized diffeomorphisms 
$\delta_\xi h_{\mu\nu}=\partial_\mu\xi_\nu+\partial_\nu\xi_\mu\,$, 
the cubic coupling ${\cal L}^{(1)}$ to a massless spin-two field $h_{\mu\nu}$ must 
arise through a bilinear conserved current of rank two, \emph{i.e.} 
$\partial_\mu\Theta^{\mu\nu}\approx 0\,$, where the weak equality
denotes the equality up to terms that vanish on the solutions of the free 
equations of motion for $\varphi_s\,$.  
For $s=\,2$, the cubic self-coupling of type $2$-$2$-$2$ coming in the Einstein--Hilbert 
Lagrangian gives rise to a conserved tensor $\Theta^{\mu\nu}$ which is equivalent 
to the Noether energy-momentum tensor $T^{\mu\nu}$ for the Fierz--Pauli  
Lagrangian. For $s\neq 2\,$, the cubic $2$-$s$-$s$ coupling ${\cal L}^{(1)}$ comes 
from the Lorentz minimal coupling prescription applied to the free Lagrangian 
${\cal L}^{(0)}$ if and only if $\Theta^{\mu\nu}$ is equal (possibly on-shell and 
modulo an ``improvement'') to the Noether energy-momentum tensor $T^{\mu\nu}$ for 
${\cal L}^{(0)}\,$. 
It is this precise condition on $\Theta^{\mu\nu}$ (for any spin!) that should be understood as ``minimal coupling'' in the Weinberg equivalence principle.

\subsubsection{Higher-order conservation laws: the higher-spin cases}

Lorentz invariance for the emission of soft massless higher ($s\geqslant 3$) 
spin particles leads to conservation laws of higher ($s-1\geqslant 2$) order, 
\textit{i.e.} for sums of products of momenta.
For generic momenta, the equation (\ref{lowen}) has no solution when $s-1>1\,$, 
therefore all coupling constants must be equal to zero: $g^{(s)}_i=0$ for any $i$ 
when $s>2\,$. In other words, as stressed by Weinberg in his book 
\cite{Weinberg:1995mt}, p.538: \textit{massless higher-spin particles may exist, 
but they cannot have couplings that survive in the limit of low energy} [that is, they cannot mediate long-range interactions]. Moreover, strictly speaking the Weinberg low-energy theorems concern only 
$s$-$s'$-$s'$ couplings. 

Nevertheless, notice the existence of a simple solution for the equation (\ref{lowen}) 
corresponding to so-called trivial scattering, \emph{i.e.} elastic scattering such 
that the outgoing particle states are permutations of the incoming ones, as in the 
case of free or possibly integrable field theories. 
For example, if we denote the ingoing momenta by $k_a$ ($a=1,2,\ldots, n$) and the 
outgoing ones by $\ell_a$, then the higher-order conservation laws 
$\sum_a g^{(s)}_a k_a^{\mu_1}\ldots k_a^{\mu_{s-1}}=(-1)^{s-1}\sum_a g^{(s)}_a \ell_a^{\mu_1}\ldots \ell_a^{\mu_{s-1}}$ of order $s-1>1$ imply that the outgoing momenta can only be permutations of the incoming ones, and that $g^{(s)}_a=g^{(s)}$ for all $a$ if $s$ is even, while $g^{(s)}_a=\epsilon_a g^{(s)}$ with $(\epsilon_a)^2=1$ for all $a$ if $s$ is odd.

\subsection{Coleman--Mandula theorem and its avatar: no higher-spin conserved charges}\label{ColMan}

The Coleman--Mandula theorem \cite{Coleman:1967ad} and its generalization to the case of supersymmetric theories with or without massless particles given by Haag, Lopuszanski and Sohnius 
\cite{Haag:1974qh} strongly restrict the symmetries of the $S$-matrix of an interacting 
relativistic field theory in four-dimensional Minkowski space-time.\footnote{For an 
extended pedagogical review, see \cite{Weinberg:2000cr}, Chapter 24.}
More precisely, (i) if the elastic two-body scattering amplitudes are generically 
non-vanishing (at almost all energies and angles); and (ii) if there is only a finite 
number of particle species on and below any given mass-shell; then the maximal 
possible extension of the Poincar\'e algebra is the (semi) direct sum of a
superalgebra (a superconformal algebra in the massless case) and an internal symmetry algebra spanned by elements that commute with the generators of the Poincar\'e algebra.

In particular, this theorem rules out higher symmetry generators (equivalently, 
conserved charges) that could have come from higher-spin symmetries surviving at 
large distances. The argument goes as follows: the gauge symmetries associated with massless particles may 
survive at spatial infinity as non-trivial rigid symmetries. In turn, such 
symmetries should lead to the conservation of some asymptotic charges. Under the 
hypotheses of the generalized Coleman--Mandula theorem, non-trivial conserved 
charges associated with asymptotic higher-spin symmetries cannot exist. 

This corollary of the generalized Coleman--Mandula theorem 
partially overlaps with the Weinberg 
low-energy theorem because the conservation law (\ref{lowen}) 
precisely corresponds to the existence of a conserved charge 
$Q^{\mu_1\ldots\,\mu_{s-1}}$ which is a symmetric tensor of rank $s-1$ that
commutes with the translations --- but does \emph{not} commute with the Lorentz 
generators.

\subsection{Generalized Weinberg--Witten theorem}

The Weinberg--Witten theorem \cite{Weinberg:1980kq} states that a 
massless particle of spin strictly greater than one \textit{cannot} possess an 
energy-momentum tensor $T_{\mu\nu}$ which is both Lorentz covariant and gauge 
invariant.\footnote{For a pedagogical essay, see e.g.
\cite{Loebbert:2008zz}.} 
Of course, this no-go theorem does not preclude gravitational interactions. 
In the spin-two case, it implies that there cannot exist any gauge-invariant 
energy-momentum tensor for the graviton. This proves that the energy of the 
gravitational field cannot be localized, but it obviously  does not prevent the 
graviton from interacting with matter or with itself. 

Recently, a refinement of the Weinberg--Witten theorem has been presented 
\cite{Porrati:2008rm} that genuinely prevents massless particles of spin 
strictly greater than \textit{two} from coupling minimally to the graviton in flat 
background. The minimality condition is stated according to the Weinberg 
equivalence principle, namely it refers to Lorentz minimal spin-two coupling 
(see Section \ref{s2case}). 
In the Lagrangian approach, the same result had already been obtained in various 
particular instances, where it had been shown that the Lorentz minimal coupling 
prescription applied to free higher-spin gauge fields enters in conflict with 
their abelian gauge symmetries 
\cite{Aragone:1979hx,Berends:1979wu,Aragone:1981yn,Boulanger:2006gr}. 
The complete no-go result ruling out the Lorentz 
minimal coupling of type $2$-$s$-$s$ in the Lagrangian approach is given in 
\cite{Boulanger:2008tg}. 

In between the Lagrangian and the $S$-matrix approaches lies the 
light-cone approach where all local cubic vertices in dimensions from four to six 
have been classified (see e.g. \cite{Metsaev:2005ar} and references therein) 
and where the same negative conclusions concerning the Lorentz minimal coupling 
of higher-spin gauge fields to gravity had already been reached and stated in 
complete generality.

This being said, consistent cubic vertices between spin-two and higher-spin gauge 
fields do exist, even in Minkowski spacetime 
\cite{Metsaev:2005ar,Boulanger:2006gr,Boulanger:2008tg}.
Instead of describing Lorentz's  minimal coupling, they contain more than two derivatives 
in total. As one can see, the generalized Weinberg--Witten theorem does not by itself 
forbid such type $2$-$s$-$s$ interactions. The crux of the matter is to combine this 
theorem with the Weinberg equivalence principle.


Together, the Weinberg equivalence principle and the generalized Weinberg--Witten
theorem do prohibit the cross-couplings of massless higher-spin particles with
low-spin particles in flat spacetime \cite{Porrati:2008rm}.
The argument goes as follows: elementary particles with spin not greater than two
are known to couple minimally to the graviton at low energy.
Therefore (Weinberg's equivalence principle) all particles interacting with
low-spin particles must also couple minimally to the graviton at low energy,
but (generalized Weinberg--Witten theorem \cite{Porrati:2008rm} and identical results
presented in \cite{Metsaev:2005ar,Boulanger:2008tg}) massless higher-spin particles
cannot couple minimally to gravity around the flat background.
Consequently, at low energies massless higher-spin particles must completely
decouple from low-spin ones. Hence, if the same Lagrangian can be used to describe
both the low-energy phenomenology and the Planck-scale physics,
then no higher-spin particles can couple to low-spin particles (including spin-2) at all.
%

\subsection{Velo--Zwanziger difficulties}

In this section, we would like to stress that, contrarily to widespread 
prejudice, the Velo--Zwanziger difficulties do not constitute a serious obstruction 
to the general programme of constructing consistent interactions involving higher-spin fields.
The observed pathologies are nothing but symptoms of non-integrability  in the sense of Cartan of the differential equations under consideration. Thus, in order to avoid pathologies,
it makes sense to follow a specific gauge 
principle\footnote{Weinberg emphasized a related point, while mentioning Velo-Zwanziger paper and other related works (c.f. refs therein), in his book 
\cite{Weinberg:1995mt}, p.244:
\textit{The problems reported with higher spin have been encountered only for higher-spin particles that have been arbitrarily assumed to have only very simple interactions with external fields. No one has shown that the problems persist for arbitrary interactions.} (...) \textit{There are good reasons to believe that the problems with higher spin disappear if the interaction with external fields is sufficiently complicated.} One may re-interpret this by stating that consistency requires less simplistic interactions, namely those governed by gauge invariance. }, which for high spins is nothing but a refined version (e.g. the Noether procedure) of the naive application of the minimal 
coupling prescription, as is the main topic of this review.

In particular, the electromagnetic interactions exhibit pathologies 
(such as seemingly superluminal propagation) in Minkowski spacetime 
already for massive spin-3/2 fields (see \cite{Velo:1970ur,Velo:1972rt} and 
a more recent analysis in \cite{Porrati:2008gv,Porrati:2008ha} which contain 
a list of other relevant references on the issue) that are therefore 
not specific to higher spins and hence deserve a separate discussion.
Indeed, the interactions between spin-$3/2$ and electromagnetic fields in 
gauged supergravities are well-known to avoid the 
Velo--Zwanziger problems.
In the case of spin-1 self-interactions, a simple model to keep in mind is the 
Born--Infeld Lagrangian, whose expansion around a non-trivial electromagnetic 
background gives a linearized theory with causal structure governed by the 
Boillat metric whose light-cone lie within that of the undeformed flat space metric --- 
see the discussion and references in \cite{Gibbons:2000xe}. 

In order to think of a model containing spins greater than one and with higher-derivative corrections that have been added following a gauge principle, one may 
immediately go to string theory, where the Born--Infeld theory is subsumed into 
open string theory. Open strings propagating in electromagnetic backgrounds 
\cite{Argyres:1989qr} contain massive spin-$s$ states with $s\geqslant \ft32$ whose 
kinetic terms contain $2s-2$ derivatives. 

The actual physical problem is how to count degrees of freedom in the presence 
of extended space-time gauge symmetries and the higher-derivative interactions that follow therefrom. In order to avoid non-integrabilities in a 
systematic fashion, a natural resolution is to abandon the standard perturbative 
approach (formulating interactions in expansions around ordinary lower-spin 
backgrounds) in favor of the unfolded approach \cite{Vasiliev:1988xc,Vasiliev:1990en,Vasiliev:1988sa,Vasiliev:1992gr}
which allows a generalized perturbative formulation of field theory in the unbroken phase as well as in various generalized metric phases and/or tensorial spacetimes \cite{Vasiliev:2001zy,Vasiliev:2001dc,Didenko:2003aa,Gelfond:2003vh}. 


To summarize this survey of no-go results, the genuine obstacles to massless 
higher-spin interactions are the Coleman--Mandula theorem, the low-energy Weinberg 
theorems, and the generalized Weinberg--Witten theorem.

\section{Possible ways out}\label{wayout}

In this Section, we discuss the weaknesses of the various hypotheses underlying the no-go theorems for interacting massless higher-spin particles in flat spacetime. Correspondingly, we present conceivable ways to surpass the spin-two barrier. Of these openings, the principal escape route is the Fradkin-Vasiliev mechanism in which the cosmological constant plays a dual role as infrared and ultraviolet regulators. This leads to Vasiliev's fully nonlinear equations, which set a new paradigm for a realm of exotic higher-spin gravities that fit naturally into the contexts of weak-weak coupling holography and tensionless limits of extended objects. This ``main route'' will be discussed in more detail in Sections 4 and 5.

\subsection{Masslessness} Implicitly, all of the aforementioned no-go theorems rely on the hypothesis of a \emph{flat} spacetime background. Indeed, the notion of massless particles is unequivocal only in theories with Poincar\'e-invariant vacua. In constantly curved non-flat spacetimes, the mass operator (\emph{i.e.} $\nabla^2$) is related to the eigenvalues of the second Casimir operators of the spacetime isometry algebra and of the Lorentz algebra. It is only in flat spacetime, however, that the eigenvalues of the mass operator are quantum numbers, which can be sent to zero leaving a strictly massless theory without any intrinsic mass-scale.  

Thus, as far as theories in Minkowski spacetime are concerned, one may consider interpreting massless higher-spin particles as limits of \textit{massive} dittos. Such particles
are consistent at low energies; on the experimental side, they are \emph{de facto} observed in hadronic physics as unstable resonances albeit not as fundamental particles\footnote{Strictly speaking, one may arguably refer to the proton as stable while already the neutron is metastable while all other massive excitations are far more short-lived.}. 
However, this high-energy limit has its own problems: it is singular in general as manifested by the van Dam--Veltman--Zakharov discontinuity in propagators of massive fields of spin greater than 3/2. Indeed, on the theoretical side, this fact is related to the complicated nature of the tensionless limit of string theory in flat spacetime.

A clear physical picture of why the high-energy limit cannot be used to find massless higher-spin particles in flat spacetime is given by the example of higher-spin resonances in quantum chromodynamics. Dimensionless quantities depend on the ratio $E/m$, where $E$ and $m$ are the energy and the mass of the resonance, respectively. As $E$ goes to infinity with $m$ kept fixed is equivalent to $m$ tending to zero keeping $E$ constant, it follows that one must send $\Lambda_{\rm QCD}$ to zero. In this limit, the size of a resonance grows indefinitely, however, and it becomes undetectable to an observer of fixed size, since the observer lives \emph{within} the resonance's Compton wavelength.\footnote{We thank one of the referees for this comment.}

\subsection{Asymptotic states and conserved charges}

The $S$-matrix theorems only concern particles that appear as asymptotic 
states. Moreover, within the perturbative approach, these asymptotic states are assumed to exist at all energy scales. Thus, an intriguing possibility is that there exists non-perturbatively defined higher-spin gauge theories in flat spacetime with mass gaps and confinement. We are not aware of any thorough investigations of such models and mechanisms so far, though Vasiliev's higher-spin gravities in four-dimensional anti-de Sitter spacetime have been conjectured to possess a perturbatively defined mass gap, resulting from dynamical symmetry breaking induced via radiative corrections \cite{Girardello:2002pp}, as we shall comment on below.

As far as confinement is concerned\footnote{This way out was briefly mentioned in the conclusions of \cite{Bekaert:2009ud}.}, one may ask whether the higher-spin charges of asymptotic states might all vanish, like for color charges in QCD. 
Incidentally, Weinberg pointed out in his book \cite{Weinberg:2000cr}, p.13, 
that some subtleties arise in the application of the Coleman--Mandula theorem in 
the presence of infrared divergences, but that \textit{there is no problem in 
non-abelian gauge theories in which all massless particles are trapped -- 
symmetries if unbroken would only govern $S$-matrix elements for gauge-neutral 
bound states}. 

\subsection{Lorentz minimal coupling}

To re-iterate slightly, the $S$-matrix no-go theorems\footnote{including  
the Coleman-Mandula theorem, since the conserved charges used in its arguments 
depend on the asymptotic behavior of interactions at large distances.} 
for higher-spin interactions are engineered for Poincar\'e-invariant relativistic quantum 
field theories aimed at describing  physics at intermediate scales lying far in between the 
Planck and Hubble scales. 
In Lagrangian terms, the generalized Weinberg-Witten theorem can essentially be understood 
as resulting from demanding compatibility between linearized gauge symmetries and the 
Lorentz minimal coupling in the absence of a cosmological constant. This compatibility requires consistent 
cubic vertices with one and two derivatives for fermions and bosons, respectively. 
Vertices with these numbers of derivatives have the same dimension as the flat-space 
kinetic terms. If consistent, they do therefore not introduce any new mass parameter. 
Hence it is natural to extrapolate the Lorentz minimal coupling to all scales. 
In doing so, however, one needs to keep in mind not only the barrier for quantum fields 
in the ultraviolet but also in the infrared.

Pertinent to this statement is the generalized Weinberg-Witten theorem. 
The assumptions are that: (i) the Lorentz minimal coupling term is always present; 
(ii) the theory extends to all energies without encountering any infrared or ultraviolet 
catastrophe. To re-iterate, the refined analysis relies crucially via assumption (i) on 
Weinberg's formulation of the equivalence principle\footnote{see Eq. (\ref{EP}) of Appendix \ref{sec:S})
or Eq. (26) in \cite{Porrati:2008rm}.}, 
which one may view as a low-energy constraint on the theory.
The result is that massless higher-spin particles cannot couple with the universal graviton 
or anything that the latter couples to. 
In other words, if such massless higher-spin theories in flat background exist in the 
mathematical sense, they \emph{cannot} be engineered to the low-energy physics that takes 
place in our Universe.

For instance, one may have a theory with two phases: A symmetric phase at high energy 
where higher-spin particles are massless and the Newton constant vanishes for all particles, 
and a broken phase, where higher-spin particles get a mass and the Newton constant is nonzero. 
This is an intriguing possibility; moreover it  probably occurs in $AdS_4$ \cite{Girardello:2002pp}, 
see the discussion below in Section \ref{break}.  
Nothing forbids  the existence of an \emph{a priori} very warm Universe where 
such exotic theories are relevant. After cooling and symmetry breakdown these may then yield 
an effective matter-coupled gravity theory in which the graviton is that field that couples 
to everything in always the same way, with a single coupling constant introduced, 
namely Newton's constant.

The assumptions (i) and (ii) are indeed vulnerable to the possibility of phase-transitions. 
This will be discussed below in Section \ref{break}. 
Looking to the limits of the experimental as well as theoretical tests of the Lorentz minimal coupling, 
there is no reason \emph{a priori} as to why the specific mechanism by which diffeomorphism invariance 
is implemented in Einstein's gravity should work at scales that are very small or very large. 
This suggests that the Lorentz minimal coupling can be rehabilitated within theories with infrared 
as well as ultraviolet cutoffs.

\subsection{Flat background}

As already stressed above, the strict 
definition of massless particle and $S$-matrix requires a flat spacetime. 
Passing to a slightly curved de Sitter or anti-de Sitter spacetime with cosmological constant $\Lambda$, one sometimes considers the existence of gauge symmetries as the criterion\footnote{This criterion is subtle, however, since for non-vanishing  $\Lambda$, generic spins cannot have as many gauge symmetries as for vanishing $\Lambda$.} of masslessness.
Since there is no genuine $S$-matrix in AdS, a subtle and fruitful way out is that 
the $S$-matrix theorems do not apply any more when the cosmological constant 
$\Lambda$ is non-vanishing, instead one ressorts to a holographic dual conformal field theory.
This way out has been exploited successfully by the Lebedev school and has given rise to cubic vertices and full nonlinear equations of motion.

\subsection{Finite-dimensionality of spacetime}

Finally, in the light of the recent progress made in amplitude calculations in 
ordinary relativistic quantum field theory \cite{Bern:2007hh,Bern:2009kd} as well 
as higher-spin gravity \cite{Giombi:2009wh,Giombi:2010vg}, one may start raising criticism against the 
very assumptions behind the Fronsdal programme: the higher-derivative nature of 
higher-spin interactions leads ultimately to a conceptual breakdown of the 
standard canonical approach to quantum field theory based on time-slicing in 
ordinary spacetime. Although one can refer perturbatively to the canonical 
structure of the free fields (thought of as fluctuations around the spin-two 
background), the non-perturbative formulation of higher-spin symmetries leads 
towards an extension of spacetime by extra bosonic coordinates on which higher-spin 
translations act by linear differentiation. One may therefore think of a 
bosonic generalization of the superspace approach to supergravities, which is 
precisely what is provided by the unfolded dynamics 
programme initiated by Vasiliev (for an illustration of the basic ideas
in the context of higher-spin supergravity, see for 
example \cite{Engquist:2002gy}).

\section{\large Various yes-go examples }\label{yesgo}

In this section we give a review of the various positive results obtained
over the years concerning consistent higher-spin cubic couplings in flat and
AdS backgrounds. Subsection \ref{flatyes} gathers together the results for
cubic vertices in flat space, while Subsection \ref{AdSyes} essentially mentions
the results obtained by Fradkin and Vasiliev in the late eighties for cubic 
vertices in (A)dS$_4\,$.
Eventually, Subsection \ref{picture} consists of a summary in the form of a 
general picture for non-abelian higher-spin gauge theory, 
which seems to emerge from the known no-go theorems and yes-go examples.
Of course, a word of caution should be added: the existence of consistent 
cubic couplings does not imply that a complete theory exists at all.
However, the existence of full interacting equations 
\cite{Vasiliev:1990en,Vasiliev:1992av,Vasiliev:2003ev} 
is a strong indication that a complete interacting Lagrangian\footnote{As a matter of fact, 
a non-standard action principle for Vasiliev's equations, which leads to a non-trivial quantization, 
was proposed in \cite{Boulanger:2011dd}.} may exist, at least in (A)dS background.
Actually, one of the open problems in higher-spin gravity is whether or not the Fronsdal 
programme can be pursued beyond the cubic order in a standard fashion. 

\subsection{Consistent cubic vertices in Minkowski spacetime}\label{flatyes}

In the eighties, the quest for high-spin interactions successfully started, 
taking flat spacetime as background.
Using the light-cone gauge approach, higher-spin $s-s'-s''$ cubic vertices in 
four space-time dimensions were found in 
\cite{Bengtsson:1983pd,Bengtsson:1983pg,Bengtsson:1986kh,Fradkin:1991iy}.
These results, in the light-cone gauge approach, were considerably generalized 
later in
\cite{Metsaev:1993gx,Metsaev:1993mj,Fradkin:1995xy,Metsaev:2005ar,Metsaev:2007rn}
with a complete classification of cubic (self- and cross-) couplings 
for arbitrary massive and massless higher-spin fields, 
bosonic and fermionic, in dimensions four, five and six. Mixed-symmetry fields
were also considered therein. 
Moreover, in \cite{Metsaev:1993ap} a wide class of cubic interactions were
obtained for arbitrary fields in arbitrary dimension.

As far as manifestly Poincar\'e-invariant vertices in the 
Lagrangian approach are concerned, Berends, Burgers and van Dam (BBvD) 
obtained a class of manifestly covariant, \emph{non-abelian} 
cubic couplings in \cite{Berends:1984rq,Berends:1984wp}. 
They used a systematization of the Noether procedure for introducing 
interactions, where the couplings are not necessarily of the form 
``gauge field times conserved current''. 
In the work \cite{Berends:1984rq}, consistent and covariant cubic couplings 
of the kind $s_1-s_2-s_2$ were obtained, for the values of $s_1$ and $s_2$ 
indicated in Table \ref{T1}. 
\begin{table}[!ht]
\centering
\begin{tabular}{c |c c c c c c c }
 ${\downarrow}_{s_1} \quad {\rightarrow}^{s_2}$ 
  & $0$  & $\frac{1}{2}$  & $1$  & $\frac{3}{2}$ &  $2$
& $\frac{5}{2}$ & $3$  \\ \hline\hline
$ 0\qquad $   & $\times$  & $\times$  & $\times$ & $\times$ & $\times$ &  &  \\
\hline
$ 1\qquad $   &$\times$  & $\times$  &$\times$ & $\times$ & $\times$ &  &  \\
\hline
$ 2\qquad $   &$\times$  & $\times$  &$\times$ & $\times$ & $\times$ & $\times$ &  
\\
\hline
$ 3\qquad $   &$\times$  & $\times$  & $\times$ & $\times$ & $\times$ & $\times$ & $\times$ \\
\hline
$n\qquad$ & $\times$
\\ \hline
\end{tabular}
\caption{\it  $s_1$-$s_2$-$s_2$ covariant vertices obtained in \cite{Berends:1984rq}.
\label{T1}}
\end{table} 
Of course, some of the vertices were already known before, 
like for example in the cases $1$-$1$-$1\,$, $2$-$2$-$2$ and 
$2$-$\frac{3}{2}$-$\frac{3}{2}$ corresponding to Yang--Mills, 
Einstein--Hilbert and ordinary supergravity theories.
There is a class of cross-interactions $s_1$-$s_2$-$s_2$ for which the cubic 
vertices could easily been written. 
This class corresponds to the ``Bell--Robinson'' line $s_1=2s_2$ 
and below this line $s_1>2s_2$ \cite{Berends:1985xx} (see \cite{Deser:1990bk} in 
the $s_1=4=2s_2$ case and some more recent considerations in 
\cite{Manvelyan:2009vy}). 
In the aforementioned region $s_1\geqslant 2 s_2\,$, 
the gauge algebra remains \emph{abelian} 
at first order in a coupling constant
although the gauge transformations for 
the spin-$s_2$ field are deformed.
The reason is that the first-order deformation of the free 
spin-$s_2$ gauge transformations involves the spin-$s_2$ field only through its 
gauge-invariant Weinberg--de Wit--Freedman field-strength 
\cite{deWit:1979pe,Weinberg:1965rz}\footnote{Note that one can trivially write 
down higher-derivative Born--Infeld-like consistent cubic interactions involving 
only gauge-invariant linearized field-strength tensors \cite{Damour:1987fp}. 
However, these interactions deform neither the gauge algebra nor the gauge
transformations at first order in some coupling constant. 
Nevertheless, they might be needed when pushing the 
non-abelian cubic vertices to the next order in the coupling constants.}. 
Although they do not lead to non-abelian gauge algebras, it is interesting that 
the cubic interactions on and below the Bell--Robinson line
(\textit{i.e.} for $s_1\geqslant 2s_2$) have the form 
``spin-$s_1$ field times current'' where the current is quadratic in 
(the derivatives of) the spin-$s_2$ field-strength 
\cite{Berends:1985xx,Deser:1990bk} and is conserved on the spin-$s_2$ shell. 
Even more interestingly, these currents can be obtained from some 
global invariances of the free theory by a Noether-like procedure, provided the 
constant parameters associated with these rigid symmetries be replaced by the 
gauge parameters of the spin-$s_1$ field (also internal color indices must be 
treated appropriately) \cite{Berends:1985xx,Deser:1990bk}.
The simplest class of cubic interactions below the Bell--Robinson line 
is provided by the couplings between 
scalar fields ($s_2=0$) and a 
collection of higher-spin tensor gauge fields through the 
Berends--Burgers--van Dam currents containing $s_1$ derivatives of the scalar 
fields \cite{Berends:1985xx}. 
Recently, they have been re-examined in 
\cite{Bekaert:2007mi,Fotopoulos:2007yq,Bekaert:2009ud} 
as a toy model for higher-spin interactions.
Notice that these cubic interactions induce, at first order in the coupling 
constant, gauge transformations for the scalar field which are non-abelian 
at second order and reproduce the group of unitary operators 
acting on free scalars on Minkowski spacetime 
\cite{Bekaert:2007mi,Bekaert:2009ud}.

As was demonstrated in \cite{Boulanger:2008tg}, in a flat background the 
non-abelian $2$-$s$-$s$ vertex is unique and involves a total number 
of $2s-2$ derivatives.
{}From $s=3$ on, the non-abelian $2$-$s$-$s$ vertex in Minkowski spacetime 
is thus ``non-minimal'' and the full Lagrangian (if any) has no chance 
of being diffeomorphism-invariant, a fact which was explicitly shown 
in \cite{Boulanger:2006gr,Boulanger:2008tg}. 
It was also shown in \cite{Boulanger:2008tg} that the unique and 
non-abelian $2$-$s$-$s$ vertex in Minkowski spacetime is nothing but the leading 
term in the flat limit of the corresponding AdS Fradkin--Vasiliev 
vertex that, among others, contains the Lorentz minimal coupling.
That the minimal Lorentz coupling term in the Fradkin-Vasiliev vertex is 
\textit{sub-leading} in the flat limit shows that the Weinberg equivalence 
principle is restored for higher-spins in AdS spacetime but is lost in the 
flat limit.
This supports the need to consider higher-spin interactions in AdS background, 
at least if one wants to make a contact between higher-spin gauge
fields and low-spin theories including Einstein--Hilbert gravity. 

Recently \cite{Bekaert:2010hp}, general results on the structure of 
cubic $s$-$s'$-$s''$ couplings ($s\leqslant s'\leqslant s''$) non-abelian 
already at this order were given, showing in 
particular that the \emph{maximum} number of derivatives involved in a non-abelian 
coupling is $2s'-1$ or $2s'-2\,$, depending on the parity of the sum 
$s+s'+s''\,$. It was also shown that the cubic vertices saturating the upper 
derivative bound have a good chance of being extended to second order in the 
deformation parameter, as far as the Jacobi identity for the gauge algebra is 
concerned.
Later on, the generic non-abelian vertices
were studied and explicitly built in \cite{Manvelyan:2010wp,Manvelyan:2010jr}. 
Some classification results were also obtained about the structure of the abelian 
cubic vertices. \textit{A posteriori}, 
the approach \cite{Manvelyan:2010wp,Manvelyan:2010jr} to the writing of covariant 
non-abelian vertices can be seen as the covariantization of the vertices already 
obtained in the light-cone approach in 
\cite{Bengtsson:1983pd,Bengtsson:1983pg,Metsaev:2005ar,Metsaev:2007rn} 
where, on top of the cubic coupling given by the light-cone gauge approach, 
terms are added which vanish in the spin-$s$ De Donder gauge. 

With the advent of string field theory in the second half of the eighties, 
the construction of higher-spin cubic vertices in flat space was carried out in 
\cite{Koh:1986vg,Bengtsson:1987jt,Cappiello:1988cd} in the so-called BRST 
approach. 
This approach was indeed motivated by the BRST first quantization of the 
string and by the tensionless limit of open string field theory. 
More recently, this analysis has been pursued in \cite{Bonelli:2003kh} and 
\cite{Buchbinder:2006eq,Fotopoulos:2007nm,Fotopoulos:2007yq} 
(a review of the last three works plus other works by the same authors 
can be found in \cite{Fotopoulos:2008ka}).
The results obtained in this framework are encouraging, for instance in the 
case of non-abelian  $s\,$-$\,0\,$-$\,0\,$ interactions \cite{Fotopoulos:2007yq}, 
although the higher-spin gauge field (self and cross) interactions 
found in \cite{Fotopoulos:2007nm} are abelian, and therefore can hardly be related 
to the non-abelian higher-spin theory of Vasiliev.

Before turning to the cubic interactions in AdS background, we would like to 
continue with our brief review of positive results for higher-spin 
cubic vertices in flat space. 
Important results have recently been obtained by analyzing the tree-level 
amplitudes of the tensile (super)string.
In what could be called a String/$S$-matrix approach, the authors of 
\cite{Polyakov:2009pk,Taronna:2010qq,Polyakov:2010qs,Sagnotti:2010at} 
obtained a plethora of vertices and recovered the vertices obtained in the 
previously cited approaches, thereby creating a direct link between 
open string theory and higher-spin gauge theory, at the dynamical level.
Moreover, in the light of the uniqueness results of \cite{Boulanger:2008tg}, 
one has a precise relation between the Fradkin--Vasiliev vertices 
and string theory. 

Generically, the idea is that the non-abelian flat space cubic vertices 
obtained in  \cite{Bekaert:2005jf,Boulanger:2008tg} (which were shown to
be related to the --- appropriately taken --- flat space limit of the 
corresponding Fradkin--Vasiliev vertices) are also the seed for the construction 
of consistent \emph{massive} higher-spin vertices in flat and AdS spacetimes.
{}From these non-abelian flat space vertices, one can systematically construct 
massive and massless vertices in AdS and flat spaces by switching on mass terms 
\`a la St\"uckelberg and cosmological constant terms.
This approach has been used with success in 
\cite{Zinoviev:2008ck,Zinoviev:2009hu}. 
See also the recent work by Zinoviev \cite{Zinoviev:2010cr} 
were the frame-like formalism for higher-spin gauge fields 
is used.  

\subsection{Cubic vertices in AdS spacetime}\label{AdSyes}

As we mentioned in the previous subsection, at cubic level (\emph{i.e.} at first 
order in perturbative deformation) Fradkin and Vasiliev found a solution to the 
higher-spin (gravitational, self and cross) interaction problem by considering 
metric perturbations around (A)dS$_4$ background
\cite{Fradkin:1987ks,Fradkin:1986qy}. 
This was later extended to five dimensions \cite{Vasiliev:2001wa},  
${\cal{N}}=1$ supersymmetry \cite{Alkalaev:2002rq} and arbitrary dimensions 
\cite{Vasiliev:2011xf}.
For a recent analysis of the Fradkin--Vasiliev mechanism in arbitrary dimension 
$D$ and in the cases $2$-$s$-$s$ and $1$-$s$-$s$, see \cite{Boulanger:2008tg}. 

The Fradkin--Vasiliev construction was the starting point of dramatic progresses 
leading recently to fully nonlinear field equations for higher-spin gauge fields 
in arbitrary dimension \cite{Vasiliev:2003ev}. We will not detail their 
construction here but we simply comment that the use of twistor 
variable and Moyal--Weyl star product is central, 
although historically the usefulness of the star product was not immediately 
recognized.
In a few words, the main problem with the higher-spin gravitational interaction 
was that, introducing the Lorentz minimal coupling terms in the action and
gauge transformations, higher-spin gauge invariance could not be satisfied any 
more. 
The solution provided by Fradkin and Vasiliev was to introduce a 
non-vanishing cosmological constant $\Lambda$ and expand the metric around 
an (A)dS background. The gauge variation of the cubic terms coming from the 
Lorentz minimal coupling around (A)dS are now canceled on the free shell, 
by the variation of a \emph{finite} tail of additional non-minimal cubic vertices, 
each of them proportional to the linearized Riemann tensor around (A)dS and 
involving more and more (A)dS-covariant derivatives compensated by 
appropriate negative powers of the cosmological constant.
In that gauge variation, the terms proportional to the free equations of motion 
are absorbed through appropriate corrections to the gauge transformations.
This solution is the \emph{Fradkin--Vasiliev mechanism}, 
and we call the gravitational cubic coupling they obtained the 
\emph{quasi-minimal coupling}, in the sense that the Lorentz minimal coupling is 
present and triggers a \emph{finite} expansion of non-minimal terms.

A salient feature of the Fradkin--Vasiliev construction is that there
are now \emph{two} independent expansion parameters. 
The AdS mass parameter $\lambda \sim \sqrt{|\Lambda|}\,$ 
and the dimensionless deformation parameter 
$g := (\lambda \ell_{\rm p})^{\frac{D-2}{2}}$  
that counts the order in the weak field expansion, where the Planck length 
$\ell_{\rm p}$ appears in front of the action through $ 1 / \ell_{\rm p}^{D-2}$ and where one
works with dimensionless physical fields. 

At the cubic level and for any given triplet of spins $\{s,s',s''\}\,$, 
there appears a finite expansion in \emph{inverse} powers of $\lambda\,$, 
where the terms with the highest negative power of $\lambda$ bring the highest 
number of (A)dS-covariant derivatives acting on the weak fields. That highest 
power of $1/\lambda\,$ is proportional to $s''\,$, so that for unbounded 
spins the Fradkin--Vasiliev cubic Lagrangian is nonlocal. 
The massive parameter $\lambda$ simultaneously (i) sets the 
\emph{infrared cutoff} via $|\Lambda|\sim\lambda^2$ and the critical masses 
$M^2 \sim\lambda^2$ for the dynamical fields; and 
(ii) dresses the derivatives in the interaction vertices thus enabling the 
Fradkin--Vasiliev mechanism. This dual r\^ole played by the cosmological constant 
is responsible for an exotic property of the Fradkin--Vasiliev cubic coupling. 
\vspace*{.4cm}

\noindent \textbf{Exotic non-locality of the Fradkin--Vasiliev Lagrangian}
\vspace*{.2cm}

\noindent In the physically relevant cases where one has a separation of length 
scales, \emph{i.e.} $\ell_{\rm p}\ll \ell \ll \l^{-1}$ where 
$\ell\sim\, \parallel\varphi\parallel/\parallel\partial\varphi\parallel$ is some
wave length characterizing the physical system under consideration and where 
$\lambda^{-1}$ denotes here a generic infrared scale, not necessarily related 
to the cosmological constant, two situations can arise for perturbatively local 
(\textit{c.f.} Subsection \ref{picture}) Lagrangians 
having vertices $V_n$ involving higher ($n\geqslant 3$) derivatives 
of the fields:

\begin{itemize}
\item[A.] \textbf{Mild non-locality}: the theory is weakly coupled 
in the sense that $V_n \sim (\ell_{\rm p}/\ell)^{n-2} \ll 1\,$. 
This situation arises for broken higher-spin symmetry, tensionful string 
sigma models etc.
\item[B.] \textbf{Exotic non-locality}: the theory is strongly coupled 
in the sense that the vertices $V_n$ are proportional to  
$(\ell\lambda)^{-n+2}\gg 1\,$. This is the situation for the Fradkin--Vasiliev 
vertices: In the derivative expansion appearing within the Fradkin--Vasiliev 
mechanism, the terms involving the maximal number of derivatives are dominant 
since they contain the infrared cutoff instead of the ultraviolet one.  
\end{itemize}

Finally, we make a comment related to the fully nonlinear Vasiliev equations
in order to show that the same behaviour appears order by order in the weak field expansion.
In this theory, the first-order corrections ${T}^{(1)}_{\mu\nu}$ 
to the stress tensor defined by $T_{\mu\nu}:=R_{\mu\nu}-\frac12 g_{\mu\nu}(R-\Lambda)$ arise in an expansion of the form 
${T}^{(1)}=\sum_{n=0}^\infty \sum_{p+q=n}\lambda^{-n}\nabla^p\varphi_s
\nabla^q\varphi_s\,$,  
see \cite{Kristiansson:2003xx} for the scalar field contributions.
One therefore sees the appearance of an \emph{infinite derivative tail} 
in the standard field equations already at first order in the weak-field 
expansion \cite{Sezgin:2002ru}. 
This would lead to tree-level amplitudes depending on the following 
two dimensionless scales: (i) the weak-field expansion coupling  
$g = (\lambda \ell_{\rm p})^{\frac{D-2}2}$ that can always be taken to be 
obey $g <\!\!\!< 1$; 
and (ii) the derivative-expansion coupling  
$(\ell\lambda )^{-n+2}$ where $\ell$ is the characteristic wavelength.
Thus the tails are strongly coupled around solutions that are close to the 
AdS$_D$ solution since here $\ell\lambda<\!\!\!< 1\,$.

\subsection{Main lessons}\label{picture}

\vspace{4mm}The first important lesson which one can draw from the previous 
discussions is that, contrarily to widespread prejudices, many doors are left 
open for massless higher-spin particles. 
The second important lesson is that interactions for higher-spin gauge fields 
exist but are rather exotic. Some of their properties clash with standard lores 
inherited from low-spin physics, and indeed, there is no fundamental reason to 
expect that higher-spin fields must behave as their low-spin companions.

Some model-independent features of non-abelian higher-spin gauge theories seem to 
emerge from all known no-go theorems and yes-go examples. 
It appears that most of the exotic properties of higher-spin fields can roughly be 
explained by mere dimensional arguments.
As we have done in the previous subsection, we introduce a parameter $\ell$ with 
the dimension of a length and rescale all objects in order to work with 
dimensionless Lagrangian $\cal L$ and fields $\varphi\,$. 
The action takes the form: 
$S=\ell^{-D}\int d^Dx\,{\cal L}(\varphi,\, \ell\,
\partial\varphi,\,\ell^2\,\partial^2\varphi, \,\ldots)$ 
where each derivative is always multiplied by a factor of $\ell\,$.
The Lagrangian counterpart of Feynman rules in $S$-matrix arguments is the weak 
field expansion, \textit{i.e.} the fields $\varphi$ are perturbations around some 
background for which the higher-spin Lagrangian $\cal L$ (if any) should admit a 
usual perturbative power expansion in terms of these fields $\varphi\,$. Around a 
stable vacuum solution, this expansion starts with a quadratic kinetic term 
${\cal L}^{(0)}$ with at most two derivatives and it goes on with vertices of 
various homogeneity degrees in $\varphi$: a cubic vertex ${\cal L}^{(1)}\,$, a 
quartic vertex ${\cal L}^{(2)}$, \textit{etc}.

In the following 
we present four general facts (of which there is no proof in full generality but 
no counter-example has ever been found) that seem to capture universal properties 
of any massless higher-spin vertex. 
\vspace*{.2cm}

\textbf{A. Higher-spin vertices are local order by order in some length scale}\label{pertloc}

A function of the field and its derivatives (treated as independent variables) is 
said to be \textit{local} if it only depends on a finite number of derivatives 
$\partial\varphi$, $\partial^2\varphi$, ...., $\partial^k\varphi$ (for some fixed 
integer $k$) and, moreover, if it only depends polynomially on these derivatives.

In the Lagrangian framework, the strong form of locality is the condition that the 
Lagrangian ${\cal L}$ must be a local function of the field $\varphi$, 
\textit{i.e.} the total number of derivatives is bounded from above (so, in our 
conventions, the Lagrangian is a polynomial in the length parameter $\ell$). 
A weaker form of locality is the requirement that the Lagrangian ${\cal L}$ is 
\emph{perturbatively local} in the sense that it admits a power series expansion in 
the fields and all their derivatives (so, in our conventions, each vertex must 
admit a power series expansion in the length scale $\ell\,$). 
Strictly speaking, this weak form of locality is rather a mild form of 
non-locality because it is obviously not equivalent to the genuine requirement of 
locality. Nevertheless, it guarantees that somehow the non-locality (if any) is 
under control: at each order in the length scale, the theory is local; the bound 
on the total number of partial derivatives is controlled by the power of $\ell\,$.
Concretely, this means that there is no strong non-locality (such as inverse 
powers of the Laplacian) and that, perturbatively, it can be treated as a local 
theory. Effective Lagrangians provide standard examples of perturbatively local 
theories.

We note in passing that, if at the cubic level one accepts to forgo the 
assumption of perturbative locality, then the higher-spin gravitational minimal 
coupling around flat space would become automatically consistent. 
Remember that, in the early attempts to 
minimally couple higher-spin particles around flat space 
\cite{Aragone:1979bm,Berends:1979wu,Aragone:1981yn}, the problem was that 
the higher-spin variation of the cubic Lagrangian creates terms 
$\delta_{\varepsilon}S^{min}\sim \int\varepsilon\cdot
(W\;\partial\varphi+\partial W\;\varphi )$ 
proportional to the spin-2 linearized Weyl tensor $W\,$, where $\varepsilon$ is the higher-spin gauge parameter.
These terms cannot be compensated by an appropriate local gauge transformation 
for the spin-2 field, since the linearized Weyl tensor (or its symmetrized and 
traceless derivative) does not vanish on-shell. However, if one accepts 
to deal with wildly nonlocal operators and inserts the formal object 
``$\Box/\Box$'' in front of the Weyl tensor, one can compensate the terms
$\int \varepsilon\cdot (\frac{1}{\Box}\,\Box W\;\partial\varphi + 
\partial \frac{1}{\Box} \Box W\;\varphi )$ by appropriate nonlocal spin-2
gauge transformations of the form $\delta h\sim \frac{1}{\Box}\,\partial^2(\varepsilon\,\partial\varphi\,+\,\partial\varepsilon\,\varphi)$, using the fact that, contrary to the Weyl tensor, 
the D'Alembertian of the Weyl tensor is proportional to the field equations
for the spin-2 field. Schematically, 
$\Box W\sim \;\partial C\,$ where ${C}$ denotes the (linearized) Cotton tensor
which is itself a linear combination of the curl of the (linearized) Einstein tensor.

\textbf{B. Higher-spin vertices are higher-derivative}\label{hder}

The higher-derivative property has been observed in all known examples of 
higher-spin cubic couplings. A summary of the general situation at 
the cubic level and in flat space is as follows:

\vspace{2mm}\noindent\textbf{Cubic interactions} \cite{Metsaev:2005ar}: 
\textit{In flat space, the total number $n$ of derivatives in any consistent 
local cubic vertex of type $s$-$s^\prime$-$s^{\prime\prime}$ (with 
$s\leqslant s^\prime\leqslant s^{\prime\prime}$) is bounded by 
$$s^\prime+s^{\prime\prime}-s\, \leqslant\, n\,\leqslant\, 
s+s^\prime+s^{\prime\prime}\,.$$
Therefore, the vertex contains at least $s^{\prime\prime}$ derivatives.}

\vspace{2mm}\noindent In other words, the value of the highest spin involved 
($s^{\prime\prime}$) gives the lowest number of derivatives that the cubic vertex 
must contain.

Notice that this proposition applies to low and higher spins. Examples of type 
$1$-$1$-$1$ and $2$-$2$-$2$ vertices are the cubic vertices in Yang-Mills and 
Einstein--Hilbert actions, they contain respectively $1$ and $2$ derivatives. 
Examples of $2$-$s$-$s$ vertices are, for low spins, the 
Lorentz minimal coupling ($s\leqslant 3/2$) where the energy-momentum tensor 
involves two derivatives (also for $s = 2$) and, for higher spins ($s>2$) 
the higher-derivative non-minimal coupling mentioned before.
The following two exotic properties of higher-spin particles are 
straightforward corollaries of results presented so far: 

\vspace{2mm}\noindent\textbf{Higher-derivative property}: \textit{In flat space, 
local cubic vertices including at least one massless particle of spin strictly 
higher than two contain three derivatives or more.}

\vspace{2mm}\noindent\textbf{Low-spin coupling}: 
\textit{In flat space, massless higher-spin particles couple non-minimally 
to low-spin particles. In (A)dS, they couple quasi-minimally, 
thereby restoring Weinberg's equivalence principle (gravitational coupling) 
and the conventional definition of electric charge (electromagnetic coupling).}
\vspace*{.2cm}

\textbf{C. Consistency requires an infinite tower of fields with unbounded spin}\label{hinf}

A local cubic vertex is said to be perturbatively consistent at second order 
if it admits a local --- possibly null --- quartic continuation such that the 
resulting Lagrangian incorporating the cubic and associated quartic vertices
(with appropriately modified gauge transformation laws) is consistent 
at second order in the perturbative coupling constant. 

Notice that the assumption of (perturbative) locality is crucial here.
If this assumption is dropped, then consistency is automatic 
beyond cubic level (see e.g. the general theorem in \cite{Barnich:1993vg}) in the sense that 
any cubic vertex can be completed by non-local quartic vertices \textit{etc}. 
It is the very assumption of (perturbative) locality that imposes very strong constraints 
on the set of possibilities.

In the local, non-abelian deformation problem, a necessary requirement for the 
consistency of cubic vertices to extend till quartic level is the closure of the 
algebra of gauge symmetries (at lowest order and possibly on-shell). 
This imposes stringent constraints on the algebra in (A)dS spacetime 
\cite{Fradkin:1986ka}: the presence of at least one higher-spin gauge field 
requires for consistency at quartic order an infinite tower of gauge fields 
with unbounded spin (more precisely the minimal spectrum seems to be a tower 
including all even spins).
At the cubic level, the coupling constants of each cubic vertex are independent 
from each other.
Another constraint coming from the consistency at quartic level is that the 
coupling constants of the cubic vertices are expressed in terms of a single one. 
Surprisingly, similar results seem to apply in Minkowski spacetime 
\cite{Metsaev:1991mt}.

When the spin is unbounded, the higher-spin interactions are non-perturbatively 
non-local but perturbatively local, in the rough sense 
that the number of derivatives is controlled by the length scale. More precisely, 
at any finite order in the power expansion in $\ell$ the vertices are local, but 
if all terms are included, as usually required for consistency at quartic level, 
then the number of derivatives is unbounded. Summarizing: 

\vspace{2mm}\noindent\textbf{Non-locality}: \textit{The number of derivatives is 
unbounded in any perturbatively local vertex including an infinite spectrum of 
massless particles with unbounded spin.}

A good news is that non-local theories do not automatically suffer from the 
higher-derivative problem. For non-local theories that are \emph{perturbatively} 
local, the problem may be treated if the free theory is well-behaved and if 
nonlocality is cured perturbatively (see \cite{Simon:1990ic} for a comprehensive 
review on this point).
\vspace*{.2cm}

\textbf{D. Massless higher-spin vertices are controlled by the infrared scale}
\label{infrared}

Concretely, in quantum field theory computations where massless particles are involved, one
makes use of infrared and ultraviolet cutoffs where $\ell_{IR}$ and $\ell_{UV}$ 
denote the corresponding length scales ($\ell_{UV}\ll\ell_{IR}$).
By definition of the cutoff prescription, the typical wavelength of physical 
excitations $\ell$ (roughly, the ``size of the laboratory'') must be such that 
$\ell_{UV}<\ell<\ell_{IR}$.

In low-spin physics, the ultraviolet scale is of the order of the Planck length: 
$\ell_{UV}\sim\ell_{\rm p}\,$, 
interactions are controlled by that ultraviolet cutoff and non-renormalizable theories 
are weakly coupled in the low energy regime $\ell\gg\ell_{\rm p}\,$. 
In higher-spin gauge theory, the situation is turned upside-down: interactions 
are controlled by the infrared cutoff $\ell_{IR\,(higher-spin)}$ 
(e.g. the AdS radius) and, since they are 
higher-derivative, the theory is strongly coupled in the high energy regime 
$\ell\ll\ell_{IR\,(higher-spin)}\,$.

\subsection{Higher-spin symmetry breakings}\label{break}

While the transition from massless to massive higher-spin particles is well 
understood at the free level via the St\"uckelberg mechanism, 
the higher-spin symmetry breaking remains deeply mysterious at 
the interacting level.
The qualitative scenario is briefly discussed in Subsection \ref{broken} and, 
finally, a tentative summary of the possible pictures is presented in 
Subsection \ref{1vsHS}.

\subsubsection*{A. Higher-spin gauge symmetries are broken at the infrared scale}
\label{broken}

At energies of the order of the infrared cutoff for the higher-spin gauge theory, 
\textit{i.e.} when $\ell\sim\ell_{IR\,(higher-spin)}$, higher-spin particles 
cannot be treated as ``massless'' any more. 
Instead, they get a mass of the order of $\ell^{-1}_{IR\,(higher-spin)}$ and, 
consequently, the higher-spin gauge symmetries are broken. 
Therefore, the no-go theorems do not apply any more. 
Hence, low-spin physics can be recovered at energy lower than the infrared cutoff 
of higher-spin gauge theory: $\ell>\ell_{IR\,(higher-spin)}\,$.

\emph{In Minkwoski spacetime}, 
a natural infrared scale of massless higher-spin particles is 
the ultraviolet scale of low-spin physics: 
$\ell_{IR\,(higher-spin)}\sim\ell_{UV\,(low-spin)}\sim\ell_{\rm p}\,$.
Then, the corresponding massive higher-spin particles have masses not smaller 
than the Planck mass and the higher-spin interactions become ``irrelevant'' 
in the low energy (sub-Planckian) regime.
By naive dimensional analysis, in the high energy (trans-Planckian) regime 
the scattering amplitudes should diverge since the theory is not (power-counting) 
renormalizable. 
However, for an infinite tower of higher-spin particles, the total scattering 
amplitudes may be extremely soft, or even finite. 
These possibilities are realized for tensile string theory around Minkowski 
spacetime where the ultraviolet scale is the string length, 
$\ell_{UV\,(string)}\sim\ell_s\,$, 
which is usually taken to be of the order of the ultraviolet scale for gravity:  
$\ell_s\sim\ell_{\rm p}\,$. 
The underlying symmetry principle behind such a phenomenon remains mysterious, 
though the standard lore is that higher-spin symmetries should play a key role 
in its understanding.

\emph{In AdS spacetime}, the situation is drastically different because the 
natural infrared scale is the radius of curvature: 
$\ell_{IR\,(higher-spin)}\sim R_{AdS}\sim \lambda^{-1}\,$
and the ultraviolet scale may remain the Planck length: 
$\ell_{UV\,(higher-spin)}\sim\ell_{\rm p}\,$. 
The high-energy limit of higher-spin gauge theory is then equivalent 
to the flat limit $\ell\ll R_{AdS}\,$. 
The Fradkin--Vasiliev cubic vertices and Vasiliev full non-linear equations 
are precisely along these lines.

\subsubsection*{B. Dynamical symmetry breaking: spin-one \textit{vs} 
higher-spin}\label{1vsHS}

The terminology ``no-go theorem'' assumes that the theorem (e.g. 
Coleman--Mandula's) is formulated negatively as the impossibility of realizing 
some idea (e.g. the mixing of internal and spacetime symmetries) under 
some conditions. 
If the idea proves to be possible then, retrospectively, the no-go theorem is read 
positively (by contraposition) as the necessity of some property (e.g. 
supersymmetry) for the idea to work. Similarly, one may speculate that maybe 
$S$-matrix no-go theorems \cite{Weinberg:1964ew,Coleman:1967ad,Porrati:2008rm} on 
massless higher-spin particles should be read positively as providing a hint (if 
not a proof) that, at the infrared scale where these theorems are valid, 
an exotic mechanism, reminiscent of mass gap and confinement in 
QCD, must necessarily take place  in any higher-spin gauge theory.
At low energy, higher-spin particles must either decouple from low-spin ones or 
acquire a mass: in both cases, asymptotic massless higher-spin states are 
unobservable.
Notice that, usually, the elusive higher-spin symmetry breaking is presented as a 
``spontaneous'' symmetry breaking like the Brout--Englert--Higgs mechanism in the 
electroweak theory, but pursuing the analogy with QCD might be fruitful and one 
could rather think of a ``dynamical'' 
symmetry breaking where the Goldstone modes would be composite fields. 
{}From holographic arguments, the authors of \cite{Girardello:2002pp} indeed 
advocated for such a scenario whereby masses for all (even) higher-spin fields in 
Vasiliev's minimal theory in AdS$_4$ are generated by quantum one-loop corrections 
while all low-spin gauge fields remain massless. 
We wish to stress the direct similarity to the Schwinger mechanism in two 
dimensional quantum electrodynamics \cite{Schwinger:1962tp} and the reminiscence 
to the saturation proposals for mass generation in three- and four-dimensional pure QCD, 
see e.g. \cite{Aguilar:2007jj,Aguilar:2010zx} and references therein. 

\vspace{2mm}A (maybe bold) way to present a summary of the two phases 
of higher-spin gauge theory is by analogy with non-abelian 
Yang-Mills theory (say quarkless QCD) whose main properties may be 
listed as follows: 

\begin{itemize}
  \item \textbf{High energy (unbroken symmetry)}: weak coupling (``asymptotic 
  freedom'')

  \item \textbf{Low energy (broken symmetry)}: strong coupling $\Longrightarrow$ 
      \textit{Non-perturbative effects} \\ All asymptotic states must be  
      massive (``mass gap'') and singlet (``color confinement'')
\end{itemize}

\noindent A plausible picture of non-abelian higher-spin gauge theory
is summarized as follows:

\begin{itemize}
	\item \textbf{High energy (unbroken symmetry)}: strong coupling
  \item \textbf{Low energy (broken symmetry)}: decoupling of massless higher-spins 
  $\Longleftarrow$ \textit{No-go theorems}\\ 
  All asymptotic higher-spin states must be massive and/or invariant under 
higher-spin symmetries
\end{itemize}

\vspace{2mm}As one can see, perhaps the biggest difficulty with non-abelian 
higher-spin gauge theory (with respect to its low-spin counterparts) is the 
absence of a phase with both unbroken symmetry \textit{and} weak coupling 
(\textit{i.e.} there is no analogue of ultraviolet ``freedom'' for Yang-Mills theory, 
or infrared ``irrelevance'' for Einstein gravity) where the theory would be easier to 
study.

\section{\large Fully interacting example: Vasiliev's higher-spin gravity}\label{Sec:VE}

After having repeated why a classically complete theory is 
key in higher-spin gravity, we lay out the salient features of 
Vasiliev's approach leading to a class of models that is not only the arguably most 
natural one but also a potentially viable brewing pot for actual semi-realistic models 
of quantum gravity. We finally address the ``state of the art'' 
and what we believe to be some ways forward.

\subsection{Examples of non-abelian gauge theories}

It is not too much of an exaggeration to stress that fact that
 \emph{the very the existence of a fully interacting non-abelian gauge field 
theory is a highly non-trivial fact, even at the classical level}. Actually, looking to 
four space-time dimensions, and focusing on bosonic gauge symmetries --- 
notwithstanding the extreme importance that supersymmetry and matter-couplings (which 
might be the same thing in higher-spin gravity) may play in order to have a 
phenomenologically viable model --- one finds essentially three classes of models 
containing local degrees of freedom: 
 
 \begin{itemize}\item Yang-Mills theories, \emph{i.e.} the theory of self-interacting 
set of spin-one fields; \item General relativity, \emph{i.e.} the theory of a 
self-interacting spin-two field;  \item Higher-spin gravity, \emph{i.e.} the theory of 
a self-interacting tower of critically massless even-spin fields. \end{itemize} 
 
Looking to their classical perturbation theories, one sees that higher-spin gravity 
distinguishes itself in the sense that it does not admit a strictly massless 
perturbative formulation on-shell in terms of massless fields in flat spacetime. 
Instead it admits a generally covariant double perturbative expansion in powers 
of\footnote{One can also define a Planck length $\ell_{\rm p}=g\sqrt{|\Lambda|}$, but 
unlike general relativity, which contains only two derivatives, higher-spin gravity has 
no sensible expansion (in its unbroken phase) in powers of $\ell_{\rm p}$. In this 
sense, the perturbation theory of higher-spin gravity is more similar in spirit to that 
of open string theory.}

\begin{itemize} 
\item a dimensionless coupling constant, $g\,$, 
counting numbers of weak fields; and 
\item the inverse of a 
 cosmological constant, $\Lambda\,$, counting numbers of pairs of derivatives. 
\end{itemize}

Although higher-spin gravity still lacks an off-shell formulation, its on-shell 
properties nonetheless suggests a  quantum theory in anti-de Sitter spacetime in which 
localized higher-spin quanta interact in such a fashion that the resulting low-energy 
effective description be dominated by higher-derivative vertices such that the standard 
minimal spin-two couplings show up only as a sub-leading term. Thus one may think of 
higher-spin gravity as an effective flat-space quantum field theory with an 
\emph{exotic cutoff}: a finite infrared cutoff, showing up as a cosmological constant 
in the gravitational perturbation theory, that at the same time plays the role of 
massive parameter in higher-derivative interactions.

Let us mention once more that the reason for this state-of-affairs can be explained 
directly in terms of the (mainly negative) results for higher-spin gauge theory in flat 
spacetime: if one removes $\Lambda$, \emph{i.e.} attempts to formulate a strictly 
massless higher-spin gauge theory without any infrared cutoff, then one falls under 
the spell of various powerful (albeit restricted) no-go theorems concerning the 
couplings between massless fields with spin $s>2$ and massless fields with spins 
$s\leqslant 2$ in flat spacetime. 

As we have already mentioned at several places, the perhaps most striking constraint on 
gauge theories with
vanishing cosmological constant, $\Lambda=0\,$, is the clear-cut clash between 
the equivalence principle, which essentially concerns the non-abelian nature of 
spin-two gauge symmetries, and abelian higher-spin gauge symmetry: on the one hand, all 
massless (as well as massive) fields must couple to a massless spin-two field via 
two-derivative vertices with the same universal coupling constant; on the other hand, 
such minimal couplings are actually incompatible with the free gauge transformations 
for spin $s>2$ fields as long as one assumes that these couplings play the dominant 
r\^ole at low energies. 

In other words, in flat spacetime there are severe no-go theorems forming a spin-two 
barrier that cannot be surpassed in the sense that massless particles of spins $s>2$ 
cannot interact with massless particles of spins $s\leqslant 2$ provided the lower-spin 
sector contains finite minimal spin-two couplings. Thus, if one wishes to proceed in 
seeking strictly massless higher-spin gauge theories (with $\Lambda=0$) then one is 
forced towards unnatural theories without any minimal spin-two couplings, whereas if 
one switches on a finite $\Lambda$ then one is naturally led into the realms of 
higher-spin gravity.

\subsection{The need for a complete theory}

Let us emphasize the need for a complete theory of higher-spin gravity already at the 
classical level, \emph{i.e.} a consistent action principle, or alternatively, set of 
equations of motion, that contains a complete set of strongly coupled derivative 
corrections.

To this end, let us return to the Fradkin--Vasiliev cancellation mechanism within the 
Fronsdal programme: in the presence of a non-vanishing cosmological constant, 
$\Lambda$, the Lorentz minimal cubic coupling (two derivatives) for a spin-$s$ field 
becomes embedded into the Fradkin--Vasiliev quasi-minimal vertex terminating in the 
non-abelian type $2$-$s$-$s$ vertex ($2s-2$ derivatives) that remains consistent in the 
$\Lambda\rightarrow 0$ limit \cite{Boulanger:2008tg} --- this ``top-vertex'' 
is thus the seed from which the subleading powers in $\Lambda$ are grown by imposing 
abelian spin-$s$ gauge invariance. The crux of the matter, however, is that the cubic 
piece of 
a complete action (consistent to all orders) may in principle contain additional 
non-minimal interactions with more derivatives that are strongly coupled in the 
$\Lambda$-expansion.

Applying dimensional analysis one arrives at the following problem: for $\Lambda<0$ the 
on-shell amplitude (Witten diagram) with three external massless gauge bosons need not 
vanish, and since $\Lambda$ now sets both the infrared cutoff (assuming the free 
theory to consist of standard tachyon-and-ghost free Fronsdal kinetic terms) and the 
mass-scale for higher-derivative vertices, the contributions to the amplitude from 
vertices with $n$ derivatives grow like the $n$th power of a large dimensionless 
number. Thus, although the top (highest-derivative) 
vertex dominates the terms with fewer derivatives inside the quasi-minimal coupling 
(including the Lorentz minimal coupling), it will in its turn be washed out by any 
genuinely non-minimal interaction, whose couplings 
(overall normalization in 
units of $\Lambda$) must hence be determined in order to estimate the three-particle 
amplitude.

Towards this end one may in principle work within a slightly refined Fronsdal programme 
as follows: (i) fix a free Fronsdal action; (ii) parameterize all consistent cubic 
vertices including \emph{a nonlocal Born--Infeld tail}, that is, a strongly coupled 
expansion in terms of Weyl tensors and their derivatives that cannot be replaced by a 
single effective Born--Infeld interaction with a finite coupling; (iii) constrain the 
spectrum and cubic couplings  by solving higher-order consistency conditions in the 
$g$-expansion (starting at quartic order). 

However, without any guiding principle other than Lorentz and gauge invariance, 
this is an \emph{a priori} intractable problem 
essentially due to the fact that the whole cubic tail must be fixed, which may require 
going to very high orders in the $g$-expansion. Of course, in the simplest scenario, 
the complete cubic action could be fixed by quartic consistency, in which case there 
would be no interaction ambiguity at the cubic level. Thus, of all possible 
hypothetical outcomes the extreme cases are: (i) quartic consistency suffices to 
completely fix the cubic action including its Born--Infeld tail; and (ii) quartic 
consistency rules out the cubic action altogether in which case the choice of free 
theory initiating the Fronsdal programme would have to be revised. 

In summary so far, to make the situation more tractable, one may resort to some 
additional guidance besides Lorentz and gauge invariance, or bias if one wishes to use 
that word, on what are suitable notions for ``higher-spin multiplets'', for selection 
of spectrum of fields, and ``higher-spin tensor calculus'', for construction of 
interactions. 

How to proceed in this issue becomes most clear in \emph{higher-spin gravity}: 
higher-spin gauge theories based on higher-spin algebras given by infinite-dimensional 
extensions of ordinary finite-dimensional space-time isometry algebras. At this stage 
it is natural to re-think how unitary representations of the complete higher-spin 
algebra are mapped directly to 
fields living in infinite-dimensional geometries containing ordinary spacetime as a 
submanifold. Indeed one of the key instruments going into Vasiliev's formulation of 
fully nonlinear equations of motion for higher-spin gravities is \emph{unfolded 
dynamics} \cite{Vasiliev:1988xc,Vasiliev:1990en,Vasiliev:1988sa,Vasiliev:1992gr}: a 
mathematically precise tool for manifestly diffeomorphism invariant generalized 
space-time reconstructions applying to finite-dimensional as well as 
infinite-dimensional cases.

\subsection{Vasiliev's equations}

A working definition of higher-spin algebras developed by Fradkin, Konstein and Vasiliev 
\cite{Fradkin:1986ka,Fradkin:1987ah,Konshtein:1988yg,Konstein:1989ij} --- that has 
proven to be useful is that of Lie subalgebras of associative algebras obtained 
from the enveloping algebras of the space-time isometry algebra by factoring out 
annihilators of their ``fundamental'', or ultra-short, unitary representations 
(singletons). In this setting, the higher-spin generators are monomials in the 
space-time isometry generators, and higher-spin multiplets arise by tensoring 
together singletons \cite{Flato:1978qz,Vasiliev:2004cm,Dolan:2005wy} which introduces 
the germ of an extended 
objects\footnote{The idea of treating algebras and their representations on a more 
equal footing --- namely as various left-, right- or two-sided modules arising 
inside the enveloping algebra and its tensor products --- is in the spirit of 
modern algebra and deformation quantization. Indeed, further development of these 
thoughts lead to first-quantized systems linking higher-spin gravities to 
tensionless strings and branes \cite{Engquist:2005yt}.} as well as a precursor to 
AdS/CFT.

In order to construct higher-spin extensions of four-dimensional gravity, the 
simplest higher-spin algebras of this type can be realized in terms of 
elementary noncommutative twistor variables. As a result the full field content 
of a special class of higher-spin gravities theories, that we can refer to as the 
minimal bosonic models and their matter-coupled and supersymmetrized extensions, 
is packed up into finite sets of ``master'' fields living on the product of 
a commutative spacetime and a noncommutative twistor space.

The feat of Vasiliev was then to realize that these master fields can be taken to 
obey remarkably simple-looking master equations built using exterior differential 
calculus on spacetime and twistor space, and star-products on twistor space, 
reproducing the standard second-order equations in perturbation theory, in about 
the same way in which Einstein's equations arise inside a set of on-shell 
superspace constraints via constraints on the torsion and Riemann two-forms.
As a result, Vasiliev's equations are diffeomorphic invariant --- in the sense of 
unfolded dynamics --- and perturbatively equivalent to a standard set of on-shell 
Fronsdal fields albeit with interactions given by a nonlocal double perturbative 
expansion resulting from the star-products. 

Looking at the twistor-space structure one sees that it services two purposes. In naive 
double perturbation theory, the expansion in the twistor variables combined with 
star-products simply generates the higher-spin tensor calculus that one may take to 
define the minimal bosonic models after which one can naively strip off all the twistor 
variables by Taylor expansion and make contact with the standard tensorial equations of 
motion after having eliminated infinite towers of auxiliary fields. 

A more careful look at these tensorial equations of motion reveals, however, 
Born--Infeld tails that are indeed strongly coupled, \emph{i.e.} formally divergent for 
ordinary localized fluctuation fields and hence inequivalent to the canonical 
Born--Infeld interactions. Focusing on classical solutions in special sectors (boundary 
conditions) one then discovers that their re-summation is tantamount to regularizations 
of star-products that requires to perform the field-theoretic calculations 
inside the twistor space, and not just by looking at Taylor expansions. 

In other words, Vasiliev's complete higher-spin gravity is essentially non-local in 
spacetime but admits a quasi-local formulation in terms of star-products on the direct 
product of commutative spacetime and non-commutative twistor space, where one can then 
proceed building classical observables and geometries for the theory. 

This somewhat awkward albeit mathematically completely well-defined situation raises 
the issue of whether Vasiliev's equations should be viewed as natural representative 
for higher-spin gravity or not? Since there are no other known examples of classes of 
higher-spin gravities with local degrees of freedom it is difficult to make any direct 
comparisons. However, lessons can be drawn by looking at the AdS/CFT correspondence. 

\subsection{AdS/CFT correspondence: Vasiliev's theory from free conformal fields}
\label{AdSCFT}

In the previous sections we have attempted to dress a dictionary between the S-matrix 
and Lagrangian approaches in the case of vanishing cosmological constant. Switching on 
the cosmological constant the notion of the S-matrix becomes deformed into that of a 
holographic conformal field theory. Thus, one way of assessing to what  
extent a higher-spin gravity is ``natural'' is to ask oneself to what extent its dual 
conformal field theory is natural.


Shortly after Maldacena's version of the AdS/CFT conjecture, which was derived within a stringy 
context involving strong/weak-coupling dual descriptions of branes, 
the question came as to what the anti-holographic dual of 
a weakly-coupled CFT could be. Since a free CFT has infinitely many conserved currents of arbitrary spin, 
in addition to the stress-energy tensor, 
it was natural to expect the AdS dual to be a higher-spin gauge theory containing a graviton. 
With a noticeable precursor \cite{Bergshoeff:1988jm}, 
such ideas emerged progressively in a series 
of papers  
\cite{HaggiMani:2000ru,Sundborg:2000wp,Konstein:2000bi,Shaynkman:2001ip,Sezgin:2001zs,
Witten2001,Mikhailov:2002bp,Sezgin:2002rt,Klebanov:2002ja,Sezgin:2003pt}: 
the idea was born in the context of the Type IIB theory on 
$AdS_5 \times S^5$ \cite{HaggiMani:2000ru,Sundborg:2000wp,Sezgin:2001zs}, 
and then pursued in a more general D-dimensional context, first at the level of kinematics 
\cite{Konstein:2000bi,Shaynkman:2001ip} and later at a dynamical level
leading to the duality conjecture between a pure bosonic higher-spin gravity in any dimension 
and a theory of (a large number of)  
free conformal scalars in the vector representation of an internal symmetry 
group \cite{Witten2001,Mikhailov:2002bp,Sezgin:2002rt}, refined to include the strongly-coupled 
fixed points of the three-dimensional $O(N)$-model and Gross-Neveu model, respectively, 
in \cite{Klebanov:2002ja} and \cite{Sezgin:2003pt}.
More precisely, the bilinear operators formed out of free fields couple to higher-spin sources 
identified as the boundary data of bulk higher-spin gauge fields.
One should stress that although the boundary CFT is quadratic, it is nevertheless non-trivial 
since the bilinear operators actually couple to background sources, therefore the bulk dual theory is 
interacting. 
The concrete relation with Vasiliev's unfolded equations in four and five dimensions was elaborated in 
\cite{Sezgin:2001zs,Sezgin:2002rt,Sezgin:2003pt}, and the fully non-linear bosonic higher-spin gravity 
in any dimension was then found in \cite{Vasiliev:2003ev}.

The agreement between Vasiliev's four-dimensional higher-spin gravity 
and the sector of bilinear operators formed out of free conformal scalars and spinors in three dimensions 
has been verified at the level of scalar cubic couplings in \cite{Petkou:2003zz,Sezgin:2003pt}, and more 
recently,
at the general cubic level in 
\cite{Giombi:2009wh,Giombi:2010vg} under certain prescriptions which still remain to be spelled out in their entirety.
Thus the question of whether Vasiliev's higher-spin gravity is natural or not is 
equivalent to the question of whether free scalars (and spinors) are natural 
building blocks for 
three-dimensional conformal field theories with (unbroken or weakly broken) higher-spin 
currents. Or put differently, thinking about Vasiliev's higher-spin gravity is about as 
natural as it is to think of three-dimensional conformal field theories starting from 
free fields. 

Intermediate developments were given in 
\cite{Leonhardt:2002sn,Das:2003vw,Leonhardt:2003du,Leigh:2003ez,Ruehl:2004kq,Bonelli:2004ve,Hartnoll:2005yc,Diaz:2006nm,Yonge:2006tn,Elitzur:2007zz}.
More recently, the full checks of the conjecture for $AdS_4/CFT_3$ at the cubic level \cite{Giombi:2009wh,Giombi:2010vg} 
prompted a revived interest in the correspondence.\footnote{Note that recently, 
in the $AdS_3/CFT_2$ framework based on the bulk theories provided in \cite{Blencowe:1988gj,Prokushkin:1998bq}, many interesting works have appeared, see e.g. 
\cite{Campoleoni:2010zq,Henneaux:2010xg,Gaberdiel:2010ar,Gaberdiel:2010pz,Castro:2010ce,
Gaberdiel:2011wb,Gaberdiel:2011zw,Gaberdiel:2011nt,Chang:2011mz,Campoleoni:2011hg,Kraus:2011ds}
and references therein.} 
For instance, the conjecture has been generalised in the presence
of a Chern-Simons gauge field on the three-dimensional boundary \cite{Aharony:2011jz,Giombi:2011kc}.
Another duality has been proposed relating bosonic Vasiliev's theory on de Sitter bulk spacetime $dS_4$ 
and fermionic scalar fields Euclidean $CFT_3$ \cite{Anninos:2011ui}.
The thermodynamic behaviour of Vasiliev's higher-spin gravity has been inferred from CFT computations 
\cite{Shenker:2011zf}.
Several attempts toward a constructive derivation of the bulk dual of a free CFT in the vector representation have been proposed, 
such as the bilocal field approach 
\cite{Das:2003vw,Koch:2010cy,Jevicki:2011ss} and the renormalisation group \cite{Douglas:2010rc}.

Here we also wish to stress that AdS/CFT is more to gauge field theory than what 
standard global-symmetry current algebra is to quantum field theory, essentially since 
the boundary currents are coupled to bulk gauge fields.
Thinking of free conformal scalar fields, the case
of two dimensions is very special, in that the stress tensor forms a closed
operator algebra (the Virasoro algebra). Indeed, already in three dimensions 
one encounters the full higher-spin current algebra as one expands the
operator product between two stress tensor generators (including a scalar
current rather than a central term). 
Thus, in the case of four-dimensional theories of quantum gravity, it seems that the 
simplest, most natural procedure, would be to start from Vasiliev-like higher-spin 
gravities and then seek symmetry breaking mechanisms that would correspond to breaking 
the higher-spin currents, followed by taking limits in which these decouple from 
operator product expansions.

In fact, by putting more emphasis on the AdS/CFT correspondence, one may provide 
further arguments \cite{Girardello:2002pp} why higher-spin gravity is a natural 
framework for seeking ultraviolet completions of general relativity.
Ordinary general relativity together with various matter couplings (and without exotic 
vertices) may then appear at low energies as the result of the dynamical higher-spin 
symmetry breaking mechanism induced by radiative corrections proposed in 
\cite{Girardello:2002pp}, provided that the induced non-critical mass-gaps grow large at 
low energies. If so, higher-spin gravity may bridge general relativity and string 
theory, which might be needed ultimately in order to achieve non-perturbative 
unitarity.

\subsection{Emergence of extended objects}
\label{sec:extended}

Let us comment briefly on the similarities and dissimilarities between 
higher-spin gravity, with its double perturbative expansion in terms of the 
dimensionless coupling $g$ and the cosmological constant $\Lambda\,$, and 
string theory, with its double perturbative expansion in terms of the 
string coupling $g_s$ and the string tension $T_s\,$. On the one hand, both of 
these theories are genuine higher-derivative theories which implies that at 
fixed orders in $g$ and $g_s\,$, respectively, there are vertices with fields 
of sufficiently high spins involving arbitrarily large inverse powers of their 
massive parameters, $\Lambda$ and $T_s\,$, respectively. Thus, in order to 
understand their respective second quantizations ($g$ and $g_s$ expansions), 
one must first obtain a sufficiently sophisticated understanding of their 
first quantizations ($\Lambda$ and $T_s$ expansions). 
Now, to its advantage string theory offers a massless window where its 
first-quantization is weakly coupled, whereas in dealing with unbroken 
higher-spin gravity one must face the whole packed-up content of its master 
fields. 

A striking similarity between open string theory and higher-spin 
gravities occurs when one considers \cite{Konstein:1989ij} extensions of the 
higher-spin algebra by an internal, associative algebra (see also \cite{Vasiliev:2004qz,Vasiliev:2005zu}). 
In such cases, there exist colored, massless spin-two fields resembling the 
spin-2 states of open strings. These states can be given Chan-Paton factors 
since their interactions are based on an associative algebra. 
This similarity was pointed out in \cite{Francia:2002pt,Francia:2006hp} 
to which we refer for related discussions. Let us note that the existence of 
colored gravitons in extended higher-spin theories does not enter in 
contradiction with the results of \cite{Boulanger:2000rq}, since there it was 
assumed that the fields considered could have spin 2 at most and the background 
was taken to be flat.

At the classical level, there remains the possibilities of having consistent 
truncations of closed string theory down to higher-spin gravity, and of 
higher-spin gravity down to general relativity. For example, both of these 
types of 
truncations may turn out to be relevant in the case of the hypothetical 
tensionless Type IIB closed string theory on $AdS_5\times S^5$ that should be 
the anti-holographic dual of free four-dimensional maximally supersymmetric 
Yang--Mills theory in its $1/N$ expansion \cite{Sundborg:2000wp,Sezgin:2002rt}. 
Here the hypothetical five-dimensional maximally supersymmetric higher-spin 
gravity (for the linearized theory see \cite{Sezgin:2001yf}) can be identified as the 
Kaluza-Klein reduction of the ``bent'' first Regge trajectory of the flat-space 
string theory \cite{Sezgin:2002rt,Bianchi:2003wx}.
The full tensionless string theory will then involve a much larger 
higher-spin symmetry algebra bringing in mixed symmetry fields with critical 
masses such that they fit into multipletons \cite{Sezgin:2002rt,Bianchi:2003wx}.
As for consistent truncations of higher-spin gravity down to possibly 
matter-coupled (super)gravities, a look at the state-of-affairs in gauged supergravities 
arising from sphere reductions \cite{deWit:1986iy,Nastase:1999cb,Cvetic:2000nc}
suggests that one should conjecture their existence in the case of maximal supersymmetry.

As far as the the Type IIB superstring is concerned, its graviton in ten-dimensional flat spacetime 
admits a deformation into a graviton of five-dimensional anti-de Sitter spacetime. 
More generally, a key physical effect of having a negative cosmological constant  
is the formation of cusps on spiky closed strings 
\cite{Gubser:2002tv,Kruczenski:2004wg} (for generalizations to membranes, see
\cite{Sezgin:2002rt}). 
At the cusps, solitonic bound states arise at the cusps, carrying the quantum numbers of 
singletons \cite{Engquist:2005yt}. In the case of folded long strings, the resulting two-singleton
closed string states are massless symmetric tensors with large spin realized \`a la Flato--Fronsdal \cite{Flato:1978qz}. In the extrapolation of this spectrum to small spins, which is tantamount to taking a tensionless limit, resides the anti-de Sitter graviton.
In \cite{Engquist:2005yt}, it was argued in  that in order for the tensionless 
limit to lead to a closed-string field theory with nontrivial interactions, it should be 
combined with sending the cosmological constant to infinity in a discretized model with fixed 
mass parameter. This yields first-quantized 0+1 dimensional models describing multi-singleton 
states. These have continuum limits given by Wess--Zumino--Witten models with gauged W-algebras 
(rather than Virasoro algebras) that can be realized in terms of symplectic bosons 
\cite{Engquist:2005yt,Engquist:2007pr} and real fermions. 

In \cite{Engquist:2005yt} it was furthermore argued that the coupling of these first-quantized 
models to higher-spin background fields requires their extension into Poisson sigma models in 
one higher dimension containing the original systems on their boundaries. In particular, in the 
case of a single singleton, that represents one string parton or membrane parton, these 
couplings are mediated via boundary and bulk vertex operators of a topological open string in 
the phase space of a singleton, that is a particular example of the C-model of 
\cite{Cattaneo:1999fm}; the consistency of this first-quantized system with disc-topology then 
requires Vasiliev's equations. 

The resulting physical picture provides a concrete realization for the germ of an extended object that is present already in the Flato--Fronsdal formula. This picture rhymes also well with the holographic framework: just as the weak-coupling stress tensor is deformed directly into the strong-coupling stress tensor on the CFT side, the graviton in higher-spin gravity is the continuation of that in closed string theory. Moreover, the fact that topological C-models underlie general associative algebras, directly explains why Vasiliev's equations are compatible with internal Chan-Paton factors. 

One is thus led to contemplate a more profound underlying framework for quantum field theory in general, based on Poisson sigma models and topological summation, and that would naturally incorporate the gauge principle as well as radiative corrections; in the case of the topological open string, the additional zero-modes arising from cutting holes in the disc may then provide a first-quantized realization of the massive Goldstone modes of the 
Girardello--Porrati--Zaffaroni mechanism \cite{Girardello:2002pp}. 

\section{\large Conclusions and Outlook}\label{Sec:conclusions}

We have discussed the key mechanism by which higher-spin gravity evades 
the no-go theorems and in particular how the equivalence principle 
is reconciled with higher-spin gauge symmetry. 

Starting in flat spacetime, massless higher-spin particles cannot be reconciled with the equivalence principle.
Nevertheless, the Weinberg--Witten theorem does not rule out 
higher-derivative energy-momentum tensors made out of higher-spin gauge fields.
Hence massless higher-spin particles may couple non-minimally to a massless spin-two particle.
However, in such case the low-energy Weinberg theorem 
would rule out the self-coupled Einstein--Hilbert 
action and minimally-coupled
matter, in particular with low spins (\emph{i.e.} $s=0$, $1/2$, $1$), in blatant contradiction
with observations.

Going to anti de Sitter spacetime, the Lorentz minimal coupling reappears but
only as a subleading term in a strongly coupled derivative expansion.
In order to do weakly coupled calculations, even at the cubic level for
higher-spin gravity, one thus needs a complete theory with the full derivative-expansion 
under control. The simplest available candidate at the moment is Vasiliev's theory.

Remarkably, not only does it resolve all the difficulties reported in the no-go 
theorems, but actually it also seems to be the simplest unbroken higher-spin gravity 
in the sense that it corresponds, via AdS/CFT, to a free conformal field theory 
with only scalar and/or fermion fields, albeit in large number. 

\vspace{2mm}

Two major open problems that need to be considered are 
\begin{itemize}
 \item \emph{Can the Fronsdal programme be pursued until quartic vertices ?}

It is not totally excluded that the answer be ``no" 
under the requirement of perturbative locality. 
Moreover, scattering amplitudes in AdS can be defined without using an action 
principle, and the recent checks of the AdS/CFT correspondence in the context 
of higher-spin gravity at the cubic level 
were done by using the unfolded formalism in the bulk theory.

\item \emph{Does the dimensionless coupling in higher-spin gravity become large 
at low energies in AdS ? }

If the answer is ``yes'' then higher-spin gravity would be a promising candidate 
for an effective quantum gravity theory. 
Drawing on our experience with QCD, 
since higher-spin gravity has been observed to be extremely soft at high energy, 
it is tempting to think that the coupling constant becomes weak in the 
ultraviolet and should grow in infrared, such that the dynamical higher spin symmetry 
breaking,
which is present already in the ultraviolet, gives rise to a finite mass gap allowing 
the identification of the low energy and low spin regime. 
 
\end{itemize}

\section*{\large Acknowledgments}

We are grateful to S. Leclercq for collaborations on several works closely
related to the present paper.
We want to thank K.~Alkalaev, G.~Barnich, A.~Bengtsson, F.~Buisseret, N.~Colombo, 
P.~P.~Cook, V.~Didenko, 
J.~Engquist, D.~Francia, M.~Henneaux, C.~Iazeolla, V.~Mathieu, K.~Meissner, R.~Metsaev, 
J.~Mourad, D.~Polyakov, M.~Porrati, A.~Sagnotti, E.~Sezgin, E.~Skvortsov, D.~Sorokin, 
Ph.~Spindel, M.~Taronna, M.~Tsulaia, M.~A.~Vasiliev, Y.~Zinoviev and Xi Yin 
for various discussions over the years.

\begin{appendix}

\section{\large Weinberg low-energy theorem: S-matrix/Lagrangian dictionary}\label{sec:Gra}

In 1964, Weinberg obtained stringent constraints on $S$-matrix elements by 
considering the effects tied to the emission of soft massless 
quanta \cite{Weinberg:1964ew}.

Consider an $S$-matrix element with $N$ external particles of momenta 
$p_i^\mu$ ($i=1,2,\ldots,N$) corresponding to the Feynman diagram
\begin{fmffile}{feynm1021}
\begin{eqnarray}
{\cal A}(p_1,\ldots,p_N)\quad =
     \parbox{50mm}{
    \begin{fmfgraph*}(60,40)
    \fmfbottomn{i}{6}
    \fmftopn{o}{6}
    \fmfblob{.15w}{b1}
    \fmflabel{$p_1$}{i1}
    \fmflabel{$p_N$}{o2}
    \fmflabel{$p_i$}{o5}
    \fmf{plain}{i2,v2,b1}
    \fmf{plain}{i3,v3,b1}
    \fmf{plain}{i4,v4,b1}
    \fmf{plain}{i5,v5,b1}
    \fmf{plain}{b1,w2,o2}
    \fmf{plain}{b1,w3,o3}
    \fmf{plain}{b1,w4,o4}
    \fmf{plain}{b1,w5,o5}
    \end{fmfgraph*}}
\label{process}
\end{eqnarray}
\end{fmffile}where all external momenta $p_i$ are on their respective mass-shells.
For the sake of simplicity, all momenta are taken to be ingoing and the 
polarizations of these particles are left implicit in $\cal A$.

\subsection{Emission of a massless particle: Lorentz \textit{versus} gauge invariances}

The amplitude for the further emission (or absorption) from any leg of a single 
massless spin-$s$ particle of momentum $q^\mu$ and polarization 
$\epsilon_{\mu_1\ldots\,\mu_s}(q)$ is denoted by
${\cal A}(p_1,\ldots,p_N;q,\epsilon)\,$: 
\vspace*{.1cm}

\begin{fmffile}{feynm2021}
\begin{eqnarray}
{\cal A}(p_1,\ldots,p_N;q,\epsilon)\; =\;
\epsilon_{\mu_1\ldots\,\mu_s}(q)\,{\cal A}^{\mu_1\ldots\mu_s}(p_1,\ldots,p_N;q)
\;=
    \parbox{50mm}{
    \begin{fmfgraph*}(60,40)
    \fmfbottomn{i}{6}
    \fmftopn{o}{6}
    \fmfblob{.15w}{b1}
    \fmflabel{$p_1$}{i1}
    \fmflabel{$p_N$}{o2}
    \fmflabel{$p_i$}{o5}
    \fmf{plain}{i2,v2,b1}
    \fmf{plain}{i3,v3,b1}
    \fmf{plain}{i4,v4,b1}
    \fmf{plain}{i5,v5,b1}
    \fmf{plain}{b1,w2,o2}
    \fmf{plain}{b1,w3,o3}
    \fmf{plain}{b1,w4,o4}
    \fmf{plain}{b1,w5,o5}
    \fmffreeze
    \fmf{photon}{w5,o6}
    \end{fmfgraph*}}
\nonumber .
\end{eqnarray}
\end{fmffile}
\vspace*{.1cm}

\noindent In general, the line of this extra particle can be attached to any 
other line, either internal or external.

In relativistic quantum field theory, the polarizations are \emph{not} 
Lorentz-covariant objects:
under Lorentz transformations, one has
$$\epsilon_{\mu_1\ldots\,\mu_s}(q)\longrightarrow 
\epsilon_{\mu_1\ldots\,\mu_s}(q)\,+\,s\,\,q_{(\mu_1}\xi_{\mu_2\ldots\,\mu_s)}(q)$$
for some symmetric tensor $\xi$ where the round bracket denotes complete 
symmetrization over the indices. This property is well-known for massless 
particles and is the counterpart of gauge invariance in the Lagrangian approach.
Lorentz-invariance of the $S$-matrix and the decoupling of spurious degrees of 
freedom thus require the condition
\begin{eqnarray}
	q_{\mu_1}{\cal A}^{\mu_1\ldots\mu_s}(p_1,\ldots,p_N;q)=0\,,\qquad 
                          \forall q  \quad.
\label{Noe}
\end{eqnarray}

\subsection{Cubic vertices}

In the particular case where the Feynman diagram (\ref{process}) is a single 
straight line, \textit{i.e.} it describes the free propagation of a single particle, 
then the modified Feynman diagram essentially is the tree-level process
\begin{fmffile}{vertex1005}
\begin{eqnarray}
\quad\quad\quad{\cal A}(p_1,p_2)\quad=\quad
    \parbox{50mm}{
    \begin{fmfgraph*}(40,40)
    \fmfbottom{i1,i2}
    \fmftop{o1,o2}
    \fmf{plain}{i1,o2}
    \fmflabel{$p_1$}{i1}
    \fmflabel{$p_2$}{o2}
    \end{fmfgraph*}}
{\cal A}(p_1,p_2;q,\epsilon)\quad=\quad
    \parbox{50mm}{
    \begin{fmfgraph*}(60,40)
    \fmfleft{i1,i2}
    \fmfright{o1,o2}
    \fmf{plain}{i1,v,i2}
    \fmf{photon}{v,o2}
    \fmflabel{$p_1$}{i1}
    \fmflabel{$p_2$}{i2}
    \fmflabel{$q$}{o2}
    \fmfdot{v}
    \end{fmfgraph*}}
\nonumber
\end{eqnarray}
\end{fmffile}
so $\Gamma^{\mu_1\ldots\mu_s}(p_1,p_2;q):={\cal A}^{\mu_1\ldots\mu_s}(p_1,p_2;q)$ 
is the part of the cubic vertex which corresponds to the Noether current in the 
Lagrangian approach. The conservation of the Noether current in the Lagrangian 
approach is equivalent to the Lorentz invariance condition (\ref{Noe}) in 
the $S$-matrix approach. 

Let us see this in more details by considering a cubic vertex of type 
$s$-$s^\prime$-$s^\prime$ with $s\neq s^\prime\,$.
The massless particle of spin $s$ is of arbitrary momentum $q^\mu$ (so off-shell) 
while the two particles of spins $s^\prime$ are on-shell with respective momenta 
$p_1$ and $p_2\,$. Writing explicitly the polarizations $\epsilon^{(1)}(p_1)$ and 
$\epsilon^{(2)}(p_2)$ of the two spin-$s^\prime\,$ particles, 
the cubic vertex takes the form 
\begin{eqnarray}
\Gamma^{\mu_1\ldots\mu_s}(p_1,p_2;q)=
\Gamma^{\mu_1\ldots\mu_s\,|\,\nu_1\ldots\nu_{s^\prime}\,|\
 \rho_1\ldots\rho_{s^\prime}}(p_1,p_2;q)
\,\epsilon^{(1)}_{\nu_1\ldots\nu_{s^\prime}}(p_1)\,
\epsilon^{(2)}_{\rho_1\ldots\rho_{s^\prime}}(p_2)\,.
\nonumber 
\end{eqnarray}
In the Lagrangian language, the cubic interaction term corresponding to 
the cubic vertex is, without loss of generality, of the form
\begin{eqnarray}
S^{(1)}[\varphi_s,\varphi_{s'}]:= \int d^Dx\; {\cal L}^{(1)}\,,\qquad 
{\cal L}^{(1)} \; :=\; \varphi_{\mu_1\ldots\mu_s} \;
\Theta^{\mu_1\ldots\mu_s}(\varphi_{s'},\varphi_{s'})
\nonumber
\end{eqnarray}
where $\Theta^{\mu_1\ldots\mu_s}$ is bilinear in $\varphi_{s^\prime}$.
More precisely, let us write the requirement of gauge invariance of the cubic 
action $S^{(1)}[\varphi_s,\varphi_{s'}]$ under linearized spin-$s$ gauge 
transformations 
$\delta_s^{(0)}\varphi_{\mu_1\ldots\mu_s} = 
        s \;\partial_{(\mu_1} \xi_{\mu_2\ldots\mu_s)}$: 
\begin{equation}
\delta_s^{(0)}S^{(1)} + \delta_s^{(1)}S^{(0)} = 0
\nonumber 
\end{equation}
where $S^{(0)}$ denotes the free part of the action, $\delta_s^{(0)}$ the free 
spin-$s$ gauge transformations and $\delta_s^{(1)}$ the gauge transformations 
taken at linear order in the fields $\{\varphi_{s'},\varphi_s\}\,$ and linear 
in the spin-$s$ gauge parameter $\xi_{\mu_1\ldots\mu_{s-1}}\,$.
The above equation implies that $\Theta^{\mu_1\ldots\mu_s}$ is a 
conserved current: 
\begin{equation}
 \partial_{\mu_1}\Theta^{\mu_1\ldots\mu_s}(\varphi_{s'},\varphi_{s'})
\approx 0
\nonumber
\end{equation}
so that the Lorentz invariance condition (\ref{Noe}) in 
the $S$-matrix approach is indeed equivalent to the conservation of the Noether current in the Lagrangian 
approach.

In momentum space,
$$
S^{(1)}= \int d^Dq\,d^Dp_1\, d^Dp_2\,{\delta(p_1+p_2+q)}\,\Gamma^{\mu_1\ldots\mu_s\,|\,\nu_1\ldots\nu_{s^\prime}\,|\,\rho_1\ldots\rho_{s^\prime}}(p_1,p_2;q)\, \varphi_{\mu_1\ldots\mu_s}(q)\,\varphi_{\nu_1\ldots\nu_{s^\prime}}(p_1)\varphi_{\rho_1\ldots\rho_{s^\prime}}(p_2)\,.
$$
{} 
The cubic vertex with the lowest number of derivatives is 
of the form
$$\Gamma^{\mu_1\ldots\mu_s\,|\,\nu_1\ldots\nu_{s^\prime}\,|\,\rho_1\ldots\rho_{s^\prime}}(p_1,p_2;q)
\propto\Gamma^{\mu_1\ldots\mu_s}(p_1,p_2;q)\eta^{\nu_1\rho_1}\ldots\eta^{\nu_{s^\prime}\rho_{s^\prime}}$$
where there is an implicit symmetrization over all $\nu$ indices and  $$\Gamma^{\mu_1\ldots\mu_s}(p_1,p_2;q)\propto(p_1-p_2)^{\mu_1}\dots(p_1-p_2)^{\mu_s}$$
is the cubic vertex for a scalar particle coupled to a spin-$s$ massless particle.
This coupling is called ``minimal'' in the sense that it contains the minimal amount of 
derivatives and also because it corresponds to a coupling with the Berends--Burgers--van 
Dam conserved currents associated with the rigid symmetries $\delta\varphi_{s'}(k)\,=\,i\,\xi^{\mu_1\ldots\mu_{s-1}}k^{\mu_1}\dots k^{\mu_{s-1}}\varphi_{s'}(k)$ \cite{Berends:1985xx} (see also \cite{Bekaert:2009ud} for more details).
In the low energy limit $q\rightarrow 0\,$, 
the only surviving cubic interaction is indeed the minimal coupling with $s$ derivatives.

The Lorentz invariance condition (\ref{Noe}) on the amplitude ${\cal A}(p_1,\ldots,p_N;q,\epsilon)\,$ for the further emission (or absorption) of a soft massless spin-$s$ particle implies
the conservation law of order $s-1$ on the $N$ external momenta
(\ref{lowen}) where each inserted minimal vertex $\Gamma^{\mu_1\ldots\mu_s}(p_i,-p_i-q;q)$ came up with a coupling constant $g^{(s)}_i$ (for more details, see e.g. \cite{Weinberg:1995mt}, Section 13.1 or \cite{Blagojevic:2002du}, Appendix G).
Equivalently, these conservation laws can be obtained from the Noether charges associated with the above-mentioned rigid symmetries.

\section{\large Weinberg--Witten theorem: a Lagrangian reformulation}
\label{sec:S}

\subsection{Weinberg--Witten theorem}

Weinberg and Witten designed their no-go theorem \cite{Weinberg:1980kq} to eliminate 
``emergent gravity'' theories where the graviton is a bound state of particles with 
spin one or lower. Its proof involves $S$-matrix manipulations which will be discussed in more details in the next subsection on its refined version.
If one assumes locality, then it becomes surprisingly easy to prove the Lagrangian version of Weinberg--Witten theorem. Let $[s]$ denote the integer part of the spin $s\,$.

\vspace{2mm}\noindent\textbf{Lemma}: \textit{Any local polynomial which is at least quadratic in a spin-$s$ massless field, non-trivial on-shell and gauge invariant, must contain at least $2\,[s]$ derivatives.}

\proof{The corollary 1 of \cite{Bekaert:2005ka} states that, on-shell, any local polynomial which is gauge invariant may depend on the gauge fields only through the Weyl-like tensors. The latter tensors contain $[s]$ derivatives thus the lemma follows.}

A straightforward corollary of this lemma is a version of Weinberg--Witten theorem.

\vspace{2mm}\noindent\textbf{Weinberg--Witten theorem} (Lagrangian formulation): 

\noindent(i) \textit{Any perturbatively local theory containing a charge current $J^\mu$ which is non-trivial, Lorentz covariant and gauge invariant, forbids massless particles of spin $s>1/2\,$.}

\noindent(ii) \textit{Any perturbatively local theory containing a Lorentz covariant and gauge invariant energy-momentum tensor $T^{\mu\nu}$ forbids massless particles of spin $s>3/2\,$.}

\proof{In the free limit, any Noether current in a perturbatively local theory must be a quadratic local polynomial. For massless fields of spin $s>1/2$, the lemma implies that this polynomial must contain at least two derivatives (or four derivatives if $s>3/2$). However, the charge current contains one derivative and the energy-momentum tensor two derivatives.}

The lower bound $s>3/2$ of this version is slightly weaker than the lower bound $s>1$ of the original Weinberg--Witten theorem \cite{Weinberg:1980kq}. Anyway the case $s=3/2$ is low-spin and thereby is not a main concern of this paper.

\subsection{Refinement of Weinberg--Witten theorem}

In \cite{Porrati:2008rm}, the author takes gauge invariance into account
in order to still use Weinberg--Witten's  argument but in a context
where the stress-energy tensor need not be gauge-invariant 
(or Lorentz-covariant, which is the same in a second-quantized setting) 
any more.

In the original work \cite{Weinberg:1980kq} a particular matrix element was 
considered: elastic scattering of a spin-$s$ massless particle off a single soft
graviton. The initial and final polarizations of the spin-$s$ particle are 
identical, say $+s\,$, its initial momentum is $p$ and its final momentum is
$p+q\,$. The graviton is \emph{off-shell} with momentum $q\,$. The matrix 
element is
\begin{eqnarray}
\langle +s, \; p+q |  \,T_{\mu\nu}\, |+s,\; p \rangle \quad . 
\label{Tmunumatrix}
\end{eqnarray}
In the soft limit $q\longrightarrow 0$ the matrix element is completely
determined by the equivalence principle, as we recalled above when reviewing
Weinberg's low energy Theorem. 
Using the relativistic normalization for one-particle states
$\langle p|p' \rangle= 2\,p_0\,(2\pi)^3\,\delta^3(\mathbf{p}-\mathbf{p'})\,$,
we get
\begin{eqnarray}
\lim_{q\rightarrow 0}
\langle +s, \; p+q |  \,T_{\mu\nu}\, |+s,\; p \rangle &=& 
p_{\mu}\,p_{\nu}\quad .
\label{EP}
\end{eqnarray}
This is tantamount to saying that, at low energy, 
the only possible coupling between gravity and everything else is done 
via the minimal coupling procedure, bringing no more than two derivatives
(or one if the spin is half-integer) in the interaction. More precisely, among all 
possible interaction terms there must always be that coming from minimal 
coupling $\partial \rightarrow \partial + \kappa\,\Gamma(h)\,$, 
with the non-vanishing coefficient $\kappa\,$ related to Newton's constant.

{Since $q$ is space-like (\emph{off-shell} soft graviton), one goes in the
frame in which $q^{\mu} = (0,-\mathbf{q}) \,$, 
$p^{\mu} = (|\mathbf{q}|/2,\mathbf{q}/2) \,$, 
$p^{\mu} + q^{\mu} = (|\mathbf{q}|/2, -\mathbf{q}/2) \,$ (the massless
spin-$s$ particle is on-shell), and deduce that
a rotation $R(\theta)$ by an angle $\theta$ around the $\mathbf{q}$ 
direction acts on the one-particle states as
$R(\theta)|p,+s\rangle = exp(\pm i\,\theta s)|p,+s\rangle\,$, 
$R(\theta)|p+q,+s\rangle = exp(\mp i\,\theta s)|p+q,+s\rangle\,$}
since $R(\theta)$ is a rotation of  $\theta$  around $\mathbf{p}$ but of
$-\theta$ around $\mathbf{p}+\mathbf{q}=-\mathbf{p}\,$. Decomposing 
$T_{\mu\nu}$ under space rotations in terms of spherical tensors as
the complex spin-zero tensor $T_{0,0}$ plus the real components
$\{T_{1,m}\}_{m=-1}^{1}\,$ and $\{T_{2,m}\}_{m=-2}^{2}\,$, one can write 
the following relation
\begin{eqnarray}
e^{\pm 2i\,\theta\,s}\langle +s, \; p+q |  T_{j,m} |+s,\; p \rangle
&=& 
\langle +s, \; p+q | R^{\dagger} T_{j,m} R|+s,\; p \rangle
\ = \ 
e^{i\,\theta\,m} \langle +s, \; p+q |  T_{j,m} |+s,\; p \rangle
\end{eqnarray}
which admits, for $s>1\,$, the only solution 
$\langle +s, \; p+q |  \,T_{\mu\nu}\, |+s,\; p \rangle = 0\,$.
Then, \emph{if $T_{\mu\nu}$ is a tensor under Lorentz transformations}
then this implies that $\langle +s, \; p+q |  \,T_{\mu\nu}\, |+s,\; p \rangle = 0\,$
in all frames, in contradiction with the equivalence principle (\ref{EP}). 
This seems to kill gravity itself, but of course in that case as it usually
happens in gauge theories, $T_{\mu\nu}$ is not a Lorentz tensor 
(which is the same as saying that $T_{\mu\nu}$ is not gauge-invariant).

One \emph{can} define matrix elements for $T_{\mu\nu}$ that transform 
as Lorentz tensors only at the price of introducing non-physical, pure-gauge 
states. This is what the author of \cite{Porrati:2008rm} did in order to 
accommodate the Weinberg--Witten argument to gauge theories for spin-$s$ fields, 
$s>1\,$ and prove that massless higher-spin particles cannot exist around
flat background if their tensor $T_{\mu\nu}$ appearing in 
$\langle +s, \; p+q |  \,T_{\mu\nu}\, |+s,\; p \rangle$ should comply with 
the equivalence principle (\ref{EP}). 

Denoting by $v$ all one-particle spin-$s$ states, whether or not
spurious (pure-gauge), the matrix element under consideration is
denoted $ \langle v',p+q|\,T_{\mu\nu}\, |v,p\rangle \,$. 
The method used in \cite{Porrati:2008rm} in order to derive the $S$-matrix
is to perform the standard perturbative expansion of the effective action
(where $g_{\mu\nu} = \eta_{\mu\nu}+\kappa\,h_{\mu\nu}$)
\begin{eqnarray}
A = \frac{1}{16\pi G}\,\int d^4x \sqrt{-g}R + \frac{1}{2}\,\int
\frac{d^4q}{(2\pi)^4}\,\widetilde{h}^*_{\mu\nu}(q) \left(
\langle v',p+q|\,T^{\mu\nu}\, |v,p\rangle + {\cal T}^{\mu\nu}
\right)+ {\cal O}(h^2)\quad .
\label{Po16}
\end{eqnarray}
The linear interaction terms include the matrix element and another effective
tensor ${\cal T}^{\mu\nu}$ which summarizes the effect of any other
matter field but that we will omit from now on without loss of generality. 
To linear order, Einstein's equations become
\begin{eqnarray}
L_{\mu\nu}^{~~~\rho\sigma}\,h_{\rho\sigma}(q) &=& 
16\pi G\,[\langle v',p+q|\,T^{\mu\nu}\, |v,p\rangle ]\quad,
\nonumber \\
L_{\mu\nu}^{~~~\rho\sigma} &=& \delta^{\rho}_{\mu}\delta_{\nu}^{\sigma}q^2 
- \eta_{\mu\nu}\eta^{\rho\sigma}q^2 -  \delta^{\rho}_{\mu}\,q_{\nu}q^{\rho}
-  \delta^{\rho}_{\nu}\,q_{\mu}q^{\rho} + \eta^{\rho\sigma}q_{\mu}q_{\nu}
+ \eta^{\mu\nu}q_{\rho}q_{\sigma}
\end{eqnarray}
which is nothing but the Fourier transform of the symmetric differential 
operator $\vec{\cal G}^{\rho\sigma}_{\mu\nu}$ acting on the spin-$2$ field
$h_{\mu\nu}$ in the linearized (in $h_{\mu\nu}$) Einstein equations 
\begin{eqnarray}
\vec{\cal G}^{\rho\sigma}_{\mu\nu}\;h_{\rho\sigma} =  \kappa \;
T_{\mu\nu}(\varphi_s,\varphi_s) + {\cal O}(\kappa^2)
\end{eqnarray}
where $T_{\mu\nu}(\varphi_s,\varphi_s)$ is the tensor bilinear in the
spin-$s$ field $\varphi_s$ that gives the
cubic $2$-$s$-$s$ vertex in the action principle 
\begin{eqnarray}
S[h_{\mu\nu},\varphi_s] &=& S^{PF}[h_{\mu\nu}] + S^{Fr}[\varphi_{s}] + 
\frac{\kappa}{2}\, \int d^Dx \;h_{\mu\nu}\,T^{\mu\nu}(\varphi_s,\varphi_s) + 
{\cal O}(\kappa^2)\;.
\end{eqnarray} 
To this same order in the metric fluctuation, a necessary condition is 
given in \cite{Porrati:2008rm} for the consistency of the gravitational 
interactions of high-spin massless particles: 
\begin{eqnarray}
\langle v,p+q|\,T^{\mu\nu}\, |v_s,p\rangle &=&  L_{\mu\nu}^{~~~\rho\sigma}\;
\Delta_{\rho\sigma}(q)
\label{Po18}
\end{eqnarray}
with $\Delta_{\rho\sigma}(q)$ analytic in a neighborhood of $q=0\,$. 

The writing (\ref{Po16}) gives to Porrati
the most general condition for the decoupling of the so-called spurious
polarization $v_s\,$ 
(that we call here sometimes ``pure-gauge'' states)
from the $S$-matrix amplitudes. 
Decoupling occurs when one can reabsorb the change in the matrix element
due to the substitution $v\rightarrow v+v_s$ with a \emph{local} field
redefinition of the graviton field. 

In the Lagrangian language, this can be seen to originate from the  
requirement of gauge invariance of the cubic action 
$S^{(1)}:= \frac{1}{2}\,\int d^Dx \;h_{\mu\nu}T^{\mu\nu}(\varphi_s,\varphi_s)$ under 
linearized gauge transformations 
\begin{eqnarray}
\delta^{(0)}h_{\mu\nu} &=& 2 \;\partial_{(\mu}
\epsilon_{\nu)}\quad,
\\
\delta^{(0)}\varphi_{\mu_1\ldots\mu_s} &=& s \;\partial_{(\mu_1}
\epsilon_{\mu_2\ldots\mu_s)}
\end{eqnarray}
up to terms that vanish on the surface of the free field
equations:
\begin{eqnarray}
\delta^{(0)}S^{(1)} + \delta^{(1)}S^{(0)} = 0 \quad
\label{canoeq} 
\end{eqnarray}
where $S^{(0)}$ denotes the free part of the action and
$\delta^{(1)}$ denotes the gauge transformations taken 
at linear order in the field $\{h,\varphi\}\,$.
The above equation can be rewritten 
\begin{eqnarray}
\int d^Dx\;\Big[\delta^{(0)}h_{\mu\nu} \;\frac{\delta S^{(1)}}{\delta 
h_{\mu\nu}}
+ \delta^{(0)}\varphi_{\mu_1\ldots\mu_s} \;
\frac{\delta S^{(1)}}{\delta \varphi_{\mu_1\ldots\mu_s} } 
+ \delta^{(1)}h^{\mu\nu} \;\vec{\cal G}^{\rho\sigma}_{\mu\nu}
\;h_{\rho\sigma}
+ \delta^{(1)}\varphi_{\mu_1\ldots\mu_s} 
\frac{\delta S^{(0)}}{\delta \varphi_{\mu_1\ldots\mu_s}}\Big] &=& 0\quad.
\nonumber
\end{eqnarray} 
If, as is assumed in the $S$-matrix approach, one takes the spin-$s$ particle
on-shell, then one sets 
$\frac{\delta S^{(0)}}{\delta \varphi_{\mu_1\ldots\mu_s}} $ to zero. 
If, in addition, one takes the Euler--Lagrange derivative of the 
result with respect to the gravitational field, noting that the only
structure for $\delta^{(1)}h_{\mu\nu}$ that can contribute to 
(\ref{canoeq}) with  
$S^{(1)} = \frac{1}{2}\,\int d^Dx \;h_{\mu\nu}T^{\mu\nu}(\varphi_s,\varphi_s)$ 
is $\delta^{(1)}h_{\mu\nu} = R_{\mu\nu}(\varphi_s,\epsilon_s)\,$, 
one finds 
\begin{eqnarray}
T_{\alpha\beta}(\varphi_s,\delta^{(0)}\varphi_s) 
+ \vec{\cal G}^{\mu\nu}_{\alpha\beta}\, R_{\mu\nu}(\varphi_s,\epsilon_s)
 &=& 0
\end{eqnarray} 
which is (up to a convention of sign in front of the Fierz--Pauli action
$S^{FP} = \frac{1}{2}\,\int h_{\mu\nu} \vec{\cal G}^{\mu\nu}_{\alpha\beta} 
h^{\alpha\beta}$)
the translation of (\ref{Po18}) in the Lagrangian language. 

Together with the principle of equivalence (\ref{EP}), the
equation (\ref{Po18}) was the main assumption of the work \cite{Porrati:2008rm}.
We see that this condition (\ref{Po18}) is derived from the main equation
(\ref{canoeq}) in the Lagrangian formalism. Apart from the assumption of
locality of $S^{(1)}$ --- which is relaxed in the $S$-matrix analysis; 
it would be interesting to see if this relaxing really gives new consistent
solutions compared to the Lagrangian analysis --- the Lagrangian analysis
of \cite{Boulanger:2006gr,Boulanger:2008tg} does not assume the equivalence
principle and is based otherwise on a weaker form of Equation (\ref{Po18}). 
That the spin-$s$ fields are put on-shell in the $S$-matrix analysis can
be viewed as an advantage (no a priori field-theoretical realization for 
the spin-$s$ fields). 
 
Based on the sole two assumptions (\ref{EP}) and (\ref{Po18}), 
Porrati is able to prove that no massless 
high-spin particle can minimally couple to gravity in flat space
in complete accordance with the previous results of 
\cite{Aragone:1979hx,Berends:1979wu,Aragone:1981yn,Metsaev:2005ar,Boulanger:2006gr} and with 
\cite{Boulanger:2008tg}.

\end{appendix}

\bibliographystyle{utphys}
\bibliography{Biblio3}

\end{document}